\documentclass[nofootinbib,twocolumn,superscriptaddress,preprintnumbers]{revtex4-1}
\usepackage[utf8]{inputenc}
\usepackage{amsfonts}
\usepackage{mathrsfs}
\usepackage{physics}
\usepackage{footnote}
\usepackage{scrextend}
\usepackage{leftidx}
\usepackage{soul}
\usepackage{graphicx}

\usepackage{color}
\definecolor{magenta}{rgb}{0.55, 0.0, 0.55}
\definecolor{orange}{rgb}{0.75, 0.25, 0.05}
\usepackage{hyperref}
\hypersetup{
     colorlinks   = true,
     citecolor    = magenta,
	 linkcolor=blue
}

\newcommand{\bfk}{ {\bf k} }

\newcommand{\GeV}{ \textrm{GeV} }

\newcommand{\Mpc}{ \textrm{Mpc} }

\newcommand{\kstar}{ k_\star }
\newcommand{\TRH}{ T_{\rm RH} }

\newcommand{\MBH}{ m_{\rm BH} }

\usepackage{amsmath,amssymb}

\begin{document}

\title{Signals of primordial black holes at gravitational wave interferometers}

\author{Jonathan Kozaczuk}
\email{jkozaczuk@physics.ucsd.edu}
\affiliation{Department of Physics, University of California, San Diego, CA 92093, USA}
\author{Tongyan Lin}
\email{tongyan@physics.ucsd.edu}
\affiliation{Department of Physics, University of California, San Diego, CA 92093, USA}
\author{Ethan Villarama}
\email{evillara@ucsd.edu}
\affiliation{Department of Physics, University of California, San Diego, CA 92093, USA}

\begin{abstract}
Primordial black holes (PBHs) can form as a result of primordial scalar perturbations at small scales. This PBH formation scenario has associated gravitational wave (GW) signatures from second-order GWs induced by the primordial curvature perturbation, and from GWs produced during an early PBH dominated era. 
We investigate the ability of next generation GW experiments, including BBO, LISA, and CE, to probe this PBH formation scenario in a wide mass range. Measuring the stochastic GW background with GW observatories can constrain the allowed parameter space of PBHs for masses $10^9$ - $10^{27}$ g. We also discuss possible GW sources from an unconstrained region where light PBHs ($<10^9$g) temporarily dominate the energy density of the universe before evaporating. We show how PBH formation impacts the reach of GW observatories to the primordial power spectrum and provide constraints implied by existing PBH bounds.
\end{abstract}

\date{\today}

\maketitle

\section{Introduction}

Primordial black holes (PBHs) may play an interesting role in astrophysics and cosmology~\cite{Carr:2020xqk,Green:2020jor}. Much attention has been paid recently to the possibility of PBHs as dark matter (DM) and as an explanation of recent LIGO/VIRGO~\cite{LIGOScientific:2020ibl} gravitational wave (GW) observations~\cite{Bird:2016dcv}. Such PBHs would need to be heavy enough to avoid Hawking evaporation on timescales comparable to the age of the Universe. Alternatively, there may have existed a significant population of light PBHs, which would have Hawking evaporated at early times. Though not corresponding to an observable PBH population today, light evaporating PBHs can be tied to other interesting phenomena, such as an early period of PBH domination~\cite{Anantua:2008am,Hooper:2019gtx}, baryogenesis~\cite{Fujita:2014hha,Morrison:2018xla,Hamada:2016jnq,Gondolo:2020uqv,Sandick:2021gew}, the production of dark radiation~\cite{Hooper:2019gtx,Fujita:2014hha, Morrison:2018xla,Lennon:2017tqq,Masina:2020xhk,Arbey:2021ysg,Gondolo:2020uqv,Sandick:2021gew}, and perhaps the production of the observed DM density, either as a product of Hawking radiation~\cite{Hooper:2019gtx,Fujita:2014hha, Morrison:2018xla,Lennon:2017tqq,Masina:2020xhk,Arbey:2021ysg} or if evaporation ends in a stable relic~\cite{MacGibbon:1987my, Barrow:1992hq, Carr:1994ar, Lehmann:2019zgt}. Such scenarios are challenging to probe. However, gravitational wave observations at interferometers can provide important insight into the formation and abundance of both evaporating and long-lived PBHs produced in the early Universe.

In this work, we will study the prospects for future GW interferometers to detect a gravitational wave background associated with PBH formation and, possibly, black hole domination, across a wide range of mass scales. The most well-studied PBH production mechanism is the collapse of primordial density perturbations from inflation. In this context, a sizable density perturbation associated with the formation of PBHs can source a stochastic GW background at second order in cosmological perturbation theory~\cite{Matarrese:1992rp,Matarrese:1993zf,Matarrese:1997ay,Noh:2004bc,Carbone:2004iv,Nakamura:2004rm,Baumann:2007zm,YUAN2021102860}. Additionally, if the density perturbations are large enough to generate a substantial population of PBHs, a period of PBH domination can ensue, which provides additional mechanisms for producing primordial GWs. Such mechanisms have been the study of recent works including \cite{Inomata:2020lmk,PhysRevD.100.043532,Domenech:2020,Domenech:2021wkk}. We will elucidate on the combined impact of these sources in what follows.

There is a large corpus of existing literature dealing with GWs associated with PBH formation. Our study complements this past work in several ways. For one, we incorporate a treatment of the effects of a possible PBH-dominated era, which provides several additional mechanisms for generating GWs and affects the observational prospects. Secondly, we study a larger range of PBH masses than is often considered, including very light PBHs which are sometimes neglected due to their evaporation at early times. Thirdly, we incorporate a state-of-the-art treatment relating the underlying inflationary perturbations to the resulting PBH mass spectrum and subsequent evolution. Finally, we present our results in a unified way alongside other observational constraints and prospects, facilitating a clear comparison of the corresponding GW interferometer reach to that of other observations sensitive to evaporating or long-lived PBHs.

Our study focuses on PBHs for which the underlying inflationary perturbations, or the effects of PBH domination, give rise to signals at GW interferometers. This occurs for sub-solar mass PBHs (assuming that the mass spectrum is approximately monochromatic), and so we will show results for masses below $\sim 10^{30}$ g. It is important to note, however, that heavier PBHs with masses above $\sim 10^{30}$ g can also induce an observable GW signal through various mechanisms. There is a vast corresponding literature covering this regime as well, especially in the context of the observed LIGO/VIRGO mergers (see e.g.~\cite{LIGOScientific:2020ibl,Clesse:2016ajp, Inomata:2016rbd}) and recent NANOGrav results~\cite{NANOGrav:2020bcs, DeLuca:2020agl, Kohri:2020qqd,PhysRevLett.126.051303}. Such scenarios can also induce GW signals at interferometers if the mass function is sufficiently extended or the corresponding inflationary perturbations approximately scale-invariant~\cite{Clesse:2018ogk}, but we will not consider this possibility further, focusing instead on PBH populations with a nearly monochromatic mass function peaked well below the solar-mass regime.

Our study is organized as follows. Section~\ref{sec:BHcosmo} discusses various aspects of primordial black hole cosmology, and introduces the relevant parameter space. Section~\ref{sec:BH_from_perturbations} details the relationship between the assumed underlying inflationary perturbations and the resulting black hole mass spectrum. Section~\ref{sec:SGWB} describes the various stochastic GW sources relevant at interferometer scales, including GWs from the formation and evaporation of PBHs, while the resulting observational prospects are discussed alongside other astrophysical and cosmological probes in Section~\ref{sec:reach}. We conclude in Section~\ref{sec:conc}. Our final results are summarized in Figs.~\ref{fig:monochromatic_bounds}-\ref{fig:sigma_bounds}.

\section{Primordial Black Hole Cosmology}
\label{sec:BHcosmo}

Primordial black holes could have meaningfully impacted our cosmic history in several ways. Depending on their mass and initial abundance, they may have led to an early period of black hole domination (BHD), injected entropy (including a possible dark radiation component~\cite{Hooper:2019gtx}) as they evaporated, or survived long enough to constitute some fraction of the observed dark matter density. In this section, we discuss these various possibilities in the context of the PBH parameter space, compared with current constraints on PBHs. 

We will describe the PBH parameter space in terms of  characteristic mass $m_{\rm BH}$ and an initial fraction of the total energy density at time $t_\star$, with
\begin{equation}
    \beta = \frac{\rho_{\textrm{BH}, \star}}{\rho_{r,\star}}.
    \label{eq:betadef}
\end{equation} 
Here $\rho_{r,\star}$ is the energy density in radiation at that time, and we will assume throughout that PBHs are formed in the early radiation-dominated FRW universe. 

\begin{figure}
    \centering
    \includegraphics[width=0.99\linewidth]{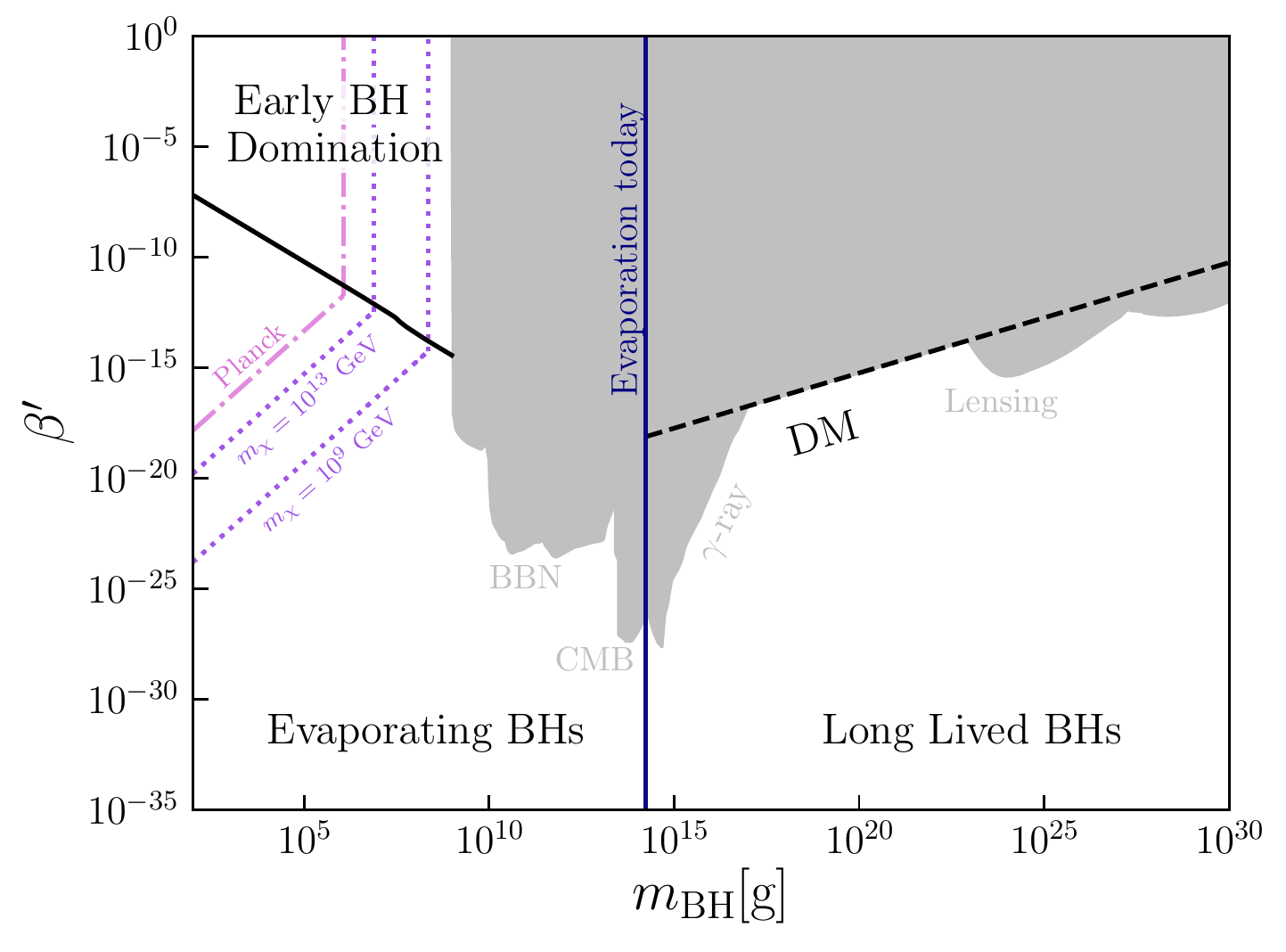}
    \caption{Range of PBH mass and $\beta'$ considered in this work. $\beta'$ is proportional to $\beta$, the initial fractional energy density in PBHs; see \eqref{eq:betaprime}. The shaded region shows observational constraints for a monochromatic mass function, see text. The solid lines delineate different regimes of PBH cosmology, while the dashed line is where PBHs comprise all of the observed DM. The dotted lines are where Hawking radiation of the PBHs into particle DM produces $\Omega_{\mathrm{DM}} h^2 \simeq 0.1$; the two lines shown are for DM mass of $10^9 \, \GeV$ and $10^{13} \, \GeV$~\cite{Hooper:2019gtx}. The dash-dotted line is where Planck-scale relics from BHs make up all of the DM~\cite{Green:1997sz}.  \label{fig:BHD_regions}} 
\end{figure}

There are several mechanisms that could in principle give rise to an abundance of PBHs, but the most well-studied is the collapse of density fluctuations from inflation, whereby  PBHs are formed when a large scalar perturbation enters the horizon. We will assume this mechanism throughout, taking a primordial curvature perturbation power spectrum that is peaked at a characteristic comoving wavenumber $k_\star$. We take $t_\star$ to be the time when the mode $k_\star$ enters the horizon, which is defined by the condition $k_\star \equiv a_\star H_\star$, with $a_\star$, $H_\star$ the FRW scale factor and Hubble parameter at horizon entry.  The PBH mass is proportional to the horizon mass at this time: 
\begin{equation}
    m_{\rm BH} = \frac{4\pi \gamma_\mathrm{eff}}{3}\rho_{r,\star} H_\star^{-3},
    \label{eq:massdef}
\end{equation}
where $\gamma_\mathrm{eff}$ is an $O(1)$ prefactor inferred from numerical simulations and dependent on the spectrum of the primordial curvature perturbation. For a review on PBH formation and discussion of this $\gamma$ parameter, see Refs. \cite{Carr:2020gox,Carr:2009jm,Green:2020jor,Gow:2020bzo}, and references therein. It is important to note that our $\gamma_{\mathrm{eff}}$ differs from the value of $\gamma \sim 0.2$ which appears sometimes in the literature. 
The difference arises from the fact that we define the PBH initial conditions at horizon entry of the mode $k_\star$, but more realistically, PBH formation happens over an extended period of time. 
In addition, the PBHs form with a distribution of masses, and here $m_\mathrm{BH}$ is defined by fitting to a lognormal mass distribution, given in  (\ref{eq:lognormal_mass_function}). Examples of the PBH mass function and details about the time-dependent formation rate will be provided in Sec.~\ref{sec:BH_from_perturbations}. 
For now, we emphasize that the value of $\gamma_{\mathrm{eff}}$ depends on the amplitude and shape of the primordial curvature perturbation.

Both $\gamma_{\rm eff}$ and $g_\star$, the number of relativistic degrees of freedom at PBH formation, will enter into a number of quantities, making it useful to define the parameter
\begin{equation}
    \beta' \equiv \sqrt{\gamma_\mathrm{eff}} \left(\frac{g_\star}{106.75}\right)^{-1/4} \left(\frac{h}{0.68}\right)^{-2} \beta,
    \label{eq:betaprime}
\end{equation}
where $h$ is the dimensionless Hubble constant. Throughout this paper, we will generally suppress $h$ dependence of quantities and use $h=0.68$. In particular, the definition above is useful is because the relic density of long-lived PBHs is proportional to $\beta'$. We thus follow the convention of plotting observational constraints in terms of $\beta'$ vs. $m_\mathrm{BH}$, and can thereby directly use the constraints of Ref.~\cite{Carr:2020gox}.

Assuming radiation domination, the temperature associated with $t_\star$ is
\begin{equation}
    T_\star = 3 \times 10^{8} \, \mathrm{GeV} \, \sqrt{\gamma_\mathrm{eff}} \left(\frac{10^{15} \mathrm{g}}{m_\mathrm{BH}} \right)^{1/2} \left(\frac{106.75}{g_\star}\right)^{1/4}.
    \label{eq:T_star}
\end{equation}
Depending on the primordial power spectrum, the actual times associated with PBH formation will be slightly different from $t_\star$. As discussed above, this leads to different characteristic PBH masses, but these effects can be captured by the $\gamma_{\rm eff}$ parameter. 
Requiring $T_\star$ to be below a reheat scale $T_{\rm RH} \sim \sqrt{H_I M_{\rm pl}}$ with the scale of inflation $H_I \lesssim 10^{14}\ \GeV$ implies that $m_\mathrm{BH} \gtrsim $ 1 g, while requiring $T_\star$ to be well above BBN temperatures implies $m_\mathrm{BH} \lesssim 10^{34}$ g.

The relationship between $m_{\rm BH}$ and $k_\star$ can be derived by redshifting back from the present day. If $\beta'$ is sufficiently small such that the PBHs never dominate the energy density, then 
\begin{align}
    \kstar & \simeq \frac{5.4 \times 10^{15}}{\Mpc} \sqrt{\gamma_\mathrm{eff}}  \left(\frac{10^{15} \mathrm{g}}{m_\mathrm{BH}}\right)^{1/2} \quad , \quad \beta' < \beta'_{\mathrm{min}},
    \label{eq:kstar_no_BHD}
\end{align}
where we have neglected the (mild) dependence on $g_\star$ in this equation. $\beta_{\mathrm{min}}$ defines the condition for black hole domination, which we will give below in \eqref{eq:BHD_line}.  

Throughout this study, we will approximate the PBH mass distribution as approximately monochromatic.
The formation scenarios we consider have sufficiently small width in the PBH mass function, such that observational constraints and cosmological evolution are expected to be modified only at the $O(1)$ level by including a full mass function. We will also neglect the effect of mergers and accretion, so that once formed, the PBHs maintain an approximately fixed mass until they begin to evaporate. Given these assumptions, the subsequent cosmological evolution of the PBH population is governed by the parameters $m_{\rm BH}$ and $\beta'$. We can therefore map the $m_{\rm BH} - \beta'$ plane into different cosmological scenarios, as shown in Fig.~\ref{fig:BHD_regions}.  In this paper, we will further map the gravitational wave reach from interferometers onto this plane.

Fig.~\ref{fig:BHD_regions} shows three qualitatively different scenarios: in the bottom right of the $m_{\rm BH}-\beta'$ plane, PBHs are long-lived enough to exist today. The region to the left of the blue vertical line features PBHs that would have evaporated by today, while the upper left corner supports an early black hole-dominated epoch. The gray shaded parameter space is excluded by existing constraints. Note that for the parameter space shown, there is (or was) at least one PBH formed in our observable universe. We discuss each of these regions in turn below.

\subsection{Long-lived PBHs}

In the bottom right portion of the parameter space in Fig.~\ref{fig:BHD_regions}, the PBHs are heavy and the initial abundance relatively small. Hawking radiation inevitably depletes the energy density of PBHs. However, the evaporation rate $d m_{\rm BH}/dt \propto 1/m_{\rm BH}^2$, so that heavy PBHs evaporate slowly. Sufficiently heavy PBHs can have lifetimes longer than the age of the Universe and exist as a (meta-)stable relic today, making up a fraction (or all) of the observed dark matter density. The evaporation time $t_{\rm evap}$ for sufficiently heavy black holes is given approximately by~\cite{Hooper:2019gtx}
\begin{equation}
t_{\rm evap} \simeq 4\times 10^{-4} {\rm s}\times \left( \frac{m_\mathrm{BH}}{10^8\,{\rm g}} \right)^3
\label{eq:t_evap}
\end{equation}
assuming only photons and neutrinos are emitted (the other SM degrees of freedom are too heavy to be thermally produced via Hawking evaporation of long-lived heavy PBHs, since the Hawking radiation temperature is inversely proportional to $m_{\rm BH}$). From this, we see that PBHs will have lifetimes longer than the age of the Universe provided $m_{\rm BH}\gtrsim 4 \times 10^{14}$g. This region lies to the right of the blue vertical line in Fig.~\ref{fig:BHD_regions}.

The long-lived population of PBHs can serve as a viable dark matter candidate. Since the effects of evaporation are negligible in this mass range, the energy density of PBHs simply redshifts as matter, $\rho_{\rm BH} \propto a^{-3}$ with $a$ the FRW scale factor.  Using entropy conservation, the PBH relic density today is given by
\begin{equation}
\Omega_{\rm BH} \approx 1.4 \times 10^{17} \beta' \left(\frac{10^{15} \mathrm{g}}{m_\mathrm{BH}} \right)^{1/2} .
\end{equation}
The parameter values where PBHs are long-lived and satisfy $ \Omega_{\rm BH}h^2 \simeq 0.12$ with $h=0.68$ is indicated by the black dashed line in Fig.~\ref{fig:BHD_regions}, and provides a particularly compelling target for GW interferometers and other astrophysical probes, as is well known (see e.g.~\cite{Cai:2018dig,Bartolo:2018rku, Bartolo:2018evs}). 
Above the DM line in Fig.~\ref{fig:BHD_regions}, the relic abundance of PBHs is too large and leads to overclosure. This is the strongest bound on the large-$\beta$ region for $m_{\rm BH} \sim 10^{17}-10^{23}$ g.

\subsection{Evaporating PBHs}

To the left of the blue vertical line in Fig.~\ref{fig:BHD_regions}, $m_{\rm BH}< 4 \times 10^{14}$g and PBHs are light enough to evaporate on timescales smaller than the age of the Universe. In this regime, PBHs themselves are no longer  a viable DM candidate, but can have other interesting effects. 
For example, in Refs.~\cite{Lennon:2017tqq,Hooper:2019gtx}, it was shown that evaporating PBHs could source a significant amount of dark radiation. The same work also pointed out that PBH evaporation can also provide a non-thermal production mechanism for heavy DM candidates, potentially giving rise to the observed DM density. The dotted lines indicate the parameters which produce $\Omega_{\rm DM} h^2 \simeq 0.1$ for particle DM mass of $10^9$ GeV (lower line) and $10^{13}$ GeV (upper line). Furthermore, it has been suggested that quantum gravity effects may halt PBH evaporation near the Planck scale, leaving behind a cold $\sim$Planck mass relic which is a possible DM candidate~\cite{Liddle:1998nt,Carr_1994}. The dash-dotted line in Fig.~\ref{fig:BHD_regions} indicates the parameters required for the Planck-scale remnants to produce $\Omega_{\rm DM} h^2 \simeq 0.1$.

\subsection{Early black hole domination \label{sec:early_BHD}}

Even if the PBHs evaporate away, if the primordial curvature perturbation is large enough, there will be a period of early black hole domination (BHD). Whether or not BHD occurs depends on if the energy density in black holes grows enough before they evaporate, so the minimum value of $\beta$ for BHD is given by the ratio $T_\mathrm{RH}/T_\star$, where the reheating temperature $T_\mathrm{RH}$ is  the temperature at the time of black hole evaporation (\ref{eq:t_evap}). Assuming matter domination, we find
\begin{equation}
    T_\mathrm{RH} = 2.8 \times 10^4\, \mathrm{GeV} \left( \frac{m_\mathrm{BH}}{10^4 \mathrm{g}} \right)^{-3/2},
\end{equation}
so our condition on the energy density for a BHD era is
\begin{equation}
    \beta'_{\mathrm{min}} \simeq 3 \times 10^{-10}  \left(\frac{10^4 \mathrm{g}}{m_\mathrm{BH}}\right).
    \label{eq:BHD_line}
\end{equation}
For simplicity, we have suppressed the dependence on degrees of freedom and on $h$. 

For $\beta' > \beta_{\rm min}'$, there is an additional BHD era where the scale factor evolves as in a matter-dominated era. This changes the relationship between the scale $k_\star$ and black hole mass $\MBH$, which now has $\beta'$ dependence:
\begin{equation}
   k_\star = \frac{2.4 \times 10^{20}}{\Mpc} \sqrt{ \gamma_{\rm eff}} \left(\frac{10^{-7}}{\beta'}\right)^{1/3} \left(\frac{10^4 \,  \mathrm{g}}{\MBH}\right)^{5/6}, \ \beta' > \beta'_\mathrm{min}
   \label{eq:kstar_BHD}
\end{equation}
since $\beta'$ determines the time of the onset of a BHD era.
In this region of parameter space, any gravitational waves generated at the time of PBH formation will be diluted by an additional redshifting factor given by
\begin{equation}
    \frac{a_\mathrm{RH}}{a_\mathrm{BHD}} =  2.4 \times 10^3 \left(\frac{\beta'}{10^{-7}}\right)^{4/3} \left(\frac{\MBH}{10^4 \, \mathrm{g}}\right)^{4/3}.
\end{equation}
This has a significant impact on experimental reach from gravitational wave interferometers if $\MBH < 10^9$ g and $\beta' > \beta'_\mathrm{min}$.

\subsection{Constraints \label{sec:constraints}}

Aside from the overclosure constraint, we show combined observational constraints on PBHs from Ref.~\cite{Carr:2020gox}. These  bounds  are  for  a  monochromatic mass function, and the observational constraints can change significantly for PBHs with extended mass functions and large enough widths~\cite{Carr:2017jsz}. In this work, we will restrict to formation scenarios where the width of the PBH mass function is not too large, such that constraints are not drastically different. We will discuss the width of the PBH mass function further in the following section.

We now briefly summarize where the constraints come from. If $10^{9}$~g$ < m_{\rm BH} < 4 \times 10^{14}$~g, then PBHs will evaporate during or after Big Bang Nucleosynthesis (BBN), but before today. This leads to injection of high energy particles that can impact BBN or the cosmic microwave background (CMB). These effects are the source of the strong constraints on $\beta$ in the parameter space for evaporating BHs. The bounds are again taken from Ref.~\cite{Carr:2020gox}, although note that the BBN bounds have been updated in more recent work~\cite{Keith:2020jww}. Bounds from observations of CMB anisotropies are strongest in the mass range of $3 \times 10^{13}$~g to $2 \times 10^{14}$~g, see for example Ref.~\cite{Stocker:2018avm}. For long-lived PBHs with mass $10^{14}$~g~$\lesssim m_{\rm BH} \lesssim  10^{17}$~g, the bounds come from various galactic and extragalactic probes of cosmic rays or gamma rays produced in the Hawking radiation. More recent work has obtained stronger constraints with data from the INTEGRAL satellite~\cite{Laha:2020ivk}.  At higher masses, there are constraints from lensing by PBHs. For $10^{23}$~g~$\lesssim m_\mathrm{BH} \lesssim 10^{28}$ g, shown here are recently updated bounds~\cite{Smyth:2019whb} based on Subaru/HSC observations~\cite{Niikura:2017zjd}. For $10^{28}$~g~$\lesssim m_\mathrm{BH} \lesssim 10^{30}$~g, OGLE bounds~\cite{Niikura:2019kqi} are shown.

\section{Primordial black holes from Curvature Perturbations} \label{sec:BH_from_perturbations}

In this work, we assume PBHs formed due to primordial curvature perturbations with Gaussian statistics. In cases with non-Gaussianity, the same curvature perturbation amplitude gives a larger energy density in PBHs~\cite{Garcia-Bellido:2017aan}. We will consider a monochromatic curvature perturbation
\begin{align}
    \mathcal{P} (k) = A \delta\left( \log (k/k_{\star}) \right),
    \label{eq:delta_perturbation}
\end{align} 
as well as a more realistic initial Gaussian perturbation
\begin{align}
    \mathcal{P} (k) =  \frac{A}{\sqrt{2 \pi \sigma ^2}} \exp\left(-\frac{\log ^2(k/k_{\star})}{2 \sigma ^2} \right).
    \label{eq:gaussian_perturbation}
\end{align}
It has been shown in \cite{Byrnes:2018txb} that the maximum primordial curvature perturbation slope is $\propto k^4$ for single-field inflation, so only a Gaussian with $\sigma \gtrsim 1$ can result in this case. Multi-field inflation \cite{PhysRevLett.125.121301,Fumagalli:2020adf,Braglia:2020taf,Braglia:2020eai} is required for growths steeper than $k^4$, which corresponds to a lognormal curvature perturbation of $\sigma<1$. We will restrict to $\sigma \le 2$ so that we can treat the black holes as forming at approximately the same time and so that we can compare with observational constraints on PBHs with nearly monochromatic mass functions.

In the section, we relate the primordial perturbations to the abundance and mass spectrum of the black holes, which can then be translated into parameters $\beta$ and $\gamma_{\rm eff}$ defined in Sec.~\ref{sec:BHcosmo}. Because PBHs are forming from rare overdensities over the vast majority of the parameter space, the relationship between the primordial curvature perturbation and the PBH mass function depends sensitively on the treatment for the collapse of density fluctuations. There is an enormous literature dedicated to this topic, with varying results. Here we consider the Press-Schechter formalism with the parameters given in~\cite{Gow:2020bzo}.  As shown in \cite{Gow:2020bzo,Young:2019osy}, this gives a relationship between BH mass spectra and curvature perturbation that is consistent with other treatments of PBH formation, namely peaks theory and variations thereof.

In the Press-Schechter formalism, PBHs are formed whenever the density contrast exceeds a certain threshold. Therefore the initial energy density can be simply related to the probability distribution for density contrasts. Consider black holes formed at a particular time corresponding to a horizon size $R$. The initial fraction of the energy density in those black holes is given by
\begin{equation}
    \beta_R = 2 \int_{C_c}^{\infty} dC \frac{m}{M_H(R)} p(C),
    \label{eq:beta_mh}
\end{equation}
where $M_H(R)$ is the horizon mass and $C$ is a smoothed density contrast, defined below. $p(C)$ is the probability distribution for $C$, and the lower limit $C_c$ is related to the density contrast threshold required for PBH formation. The black hole mass is related to the density contrast and horizon mass at formation by the critical collapse scaling relation:
\begin{equation}
    m = M_H(R) K (C - C_c)^\gamma,
\end{equation}
where $K$ and $\gamma$ are constants determined via numerical simulation.  Here we use $K=10, C_c = 2.5,$ and $\gamma = 0.36$. The critical collapse relationship allows us to recast the integrand of (\ref{eq:beta_mh}) as a function of black hole mass $m$, which gives us the mass function of black holes formed at a certain horizon size, $d \beta_R/dm$. 

To relate the curvature power spectrum in Fourier space to density contrasts in real space, we must smooth the power spectrum on horizon scales $R$ with window function $W(k,R)$. The variance of density contrasts on a scale $R$ is given by
\begin{equation}
    \sigma_0^2 = \int_0^\infty \frac{dk}{k} \frac{16}{81} (kR)^4 W^2 (k, R) \mathcal{P}(k).
    \label{eq:0th_moment}
\end{equation}
In Press-Schechter, the probability density for density contrasts is given in terms of this variance:
\begin{equation}
    p(C) = \frac{1}{\sqrt{2\pi} \sigma_0} e^{-\frac{C^2}{2 \sigma_0^2}}.
    \label{eq:pC}
\end{equation}
Note that the probability density $p(C)$ depends on the horizon size, which appears in the calculation of $\sigma_0$ through the smoothing of the primordial power spectrum.  It has been noted that different choices of window functions can give rise to quite different required amplitudes for PBH formation \cite{PhysRevD.97.103528}, but \cite{Gow:2020bzo} resolves these issues and shows that these differences are $<10 \%$ if one uses consistent quantities for smoothing and threshold density contrasts. In these calculations, we use the Gaussian window function.

The total effective energy density fraction in black holes is obtained by integrating (\ref{eq:beta_mh}) over PBHs formed at all horizon sizes, multiplied by a redshifting factor to account for the fact the black holes are forming in a radiation dominated era
\begin{equation}
    \beta(t) = \int_0^\infty \frac{dR}{R} \frac{R(t)}{R} \beta_R,
    \label{eq:beta_time}
\end{equation}
where $R$ is the horizon size. Note that this is slightly different from the true energy density at $t$, since integrating over all horizon sizes includes black holes forming at times greater than $t$. This is a negligible difference when evaluated at $t_\star$, since most black holes form before $t_\star$. Recall that the $\beta$ parameter introduced in \eqref{eq:betadef} is the effective PBH energy density evaluated at the horizon entry for the mode $k_\star$, and thus corresponds to $\beta(t_\star) \equiv \beta_\star$. Evaluating $\beta_\star$ from \eqref{eq:beta_time}, we find that the relationship to $A$ is well-described by the following fitting function:
\begin{equation}
    \beta_\star = a A^{b} \mathrm{Erfc} \left[\frac{c}{\sqrt{A}} \right],
\end{equation}
where the fit parameters $a, b, c$ will vary with the curvature perturbation width. Concretely, $a=5.47,b=0.51,c=0.28$ for the monochromatic case; $a=14.3,b=0.66,c=0.47$ for a Gaussian  perturbation with $\sigma=1$; and $a=36.0,b=0.91,c=0.64$ for a Gaussian perturbation with $\sigma=2$.

\begin{figure}
    \includegraphics[width=0.98\linewidth]{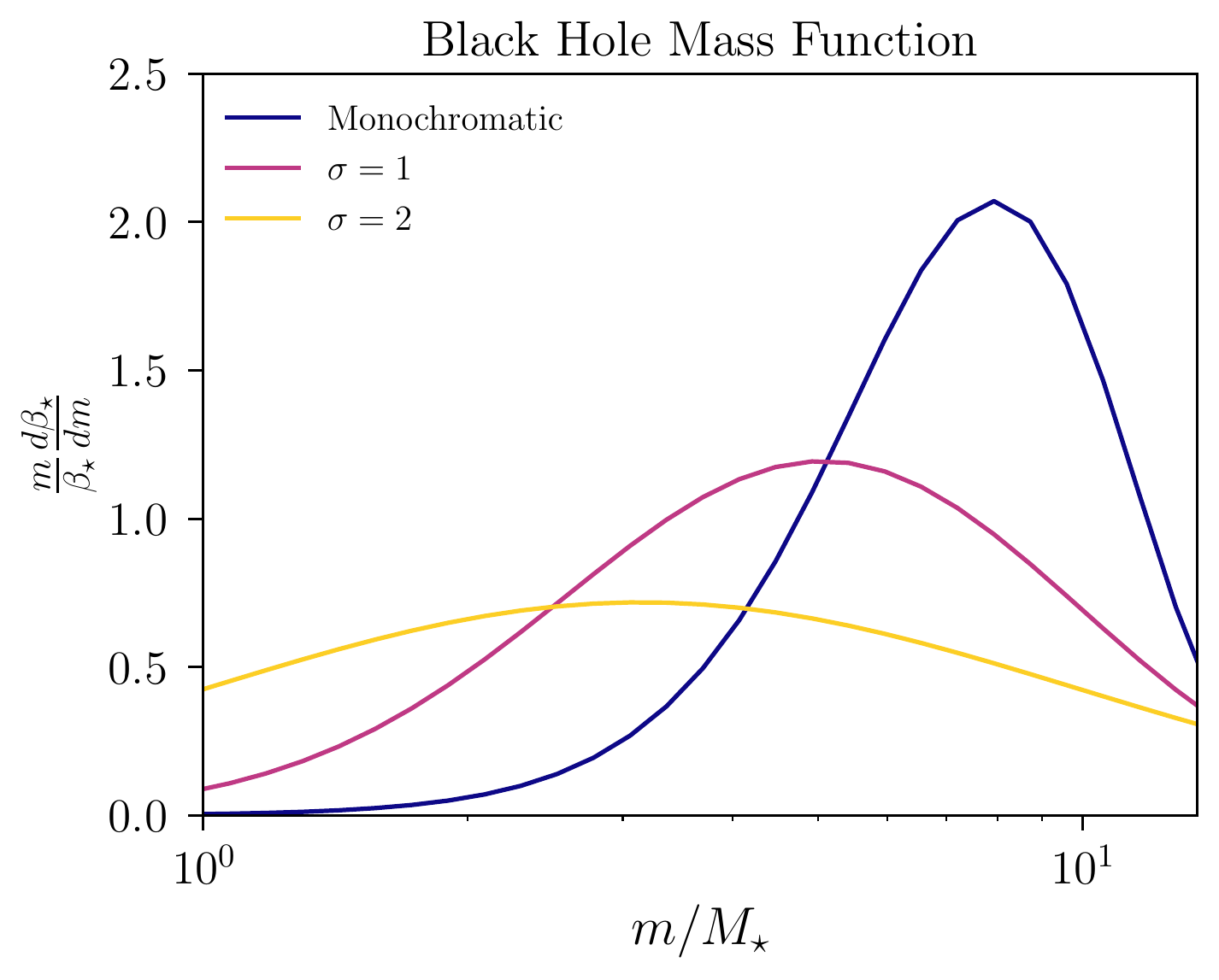}
    \caption{PBH mass function for different primordial curvature spectrum: the $\sigma = 1,2$ lines are for the spectrum in (\ref{eq:gaussian_perturbation}). The amplitude is selected so that $\beta_\star \simeq 4 \times 10^{-7}$ for all cases here.  $M_\star$ is the horizon mass associated with entry of the mode $k_\star$. Fitting these to log-normal mass functions gives peak BH masses of $m_{\rm BH} = \gamma_{\rm eff} M_\star$ with $\gamma_{\rm eff} \approx 7.6, 5.0, 3.1 $ and widths $\sigma_m \approx 0.4, 0.7, 1.1$ for the monochromatic, $\sigma = 1$, and $\sigma=2$ cases, respectively. }
    \label{fig:mass_function}
\end{figure}

The PBH mass function is given by 
\begin{equation}
\frac{1}{\beta(t)}\frac{d \beta(t)}{d m}.
\label{eq:mass_spectrum}
\end{equation}
and can be obtained by writing $\beta_R$ as an integral over $d \beta_R/dm$ in (\ref{eq:beta_time}).
Examples of the resulting mass functions are shown in Fig.~\ref{fig:mass_function}. In order to compare to existing observational bounds on PBHs with extended mass functions, we fit this mass distribution to a log-normal mass function given by
\begin{equation}
    \frac{1}{\sqrt{2\pi} \sigma_m m} \exp \left( - \frac{\log^2 (m/m_\mathrm{BH})}{2\sigma_m^2} \right).
    \label{eq:lognormal_mass_function}
\end{equation}
The fit value $m_\mathrm{BH}$ is the characteristic black hole mass. $m_\mathrm{BH}$ can be written as some pre-factor $\gamma_\mathrm{eff}(\sigma,A)$ times $M_\star $ where $M_\star = 4\pi \rho_{r,\star} H_\star^{-3}/3$, the horizon mass corresponding to the entry of the mode $k_\star$.

\begin{figure}
    \includegraphics[width=0.98\linewidth]{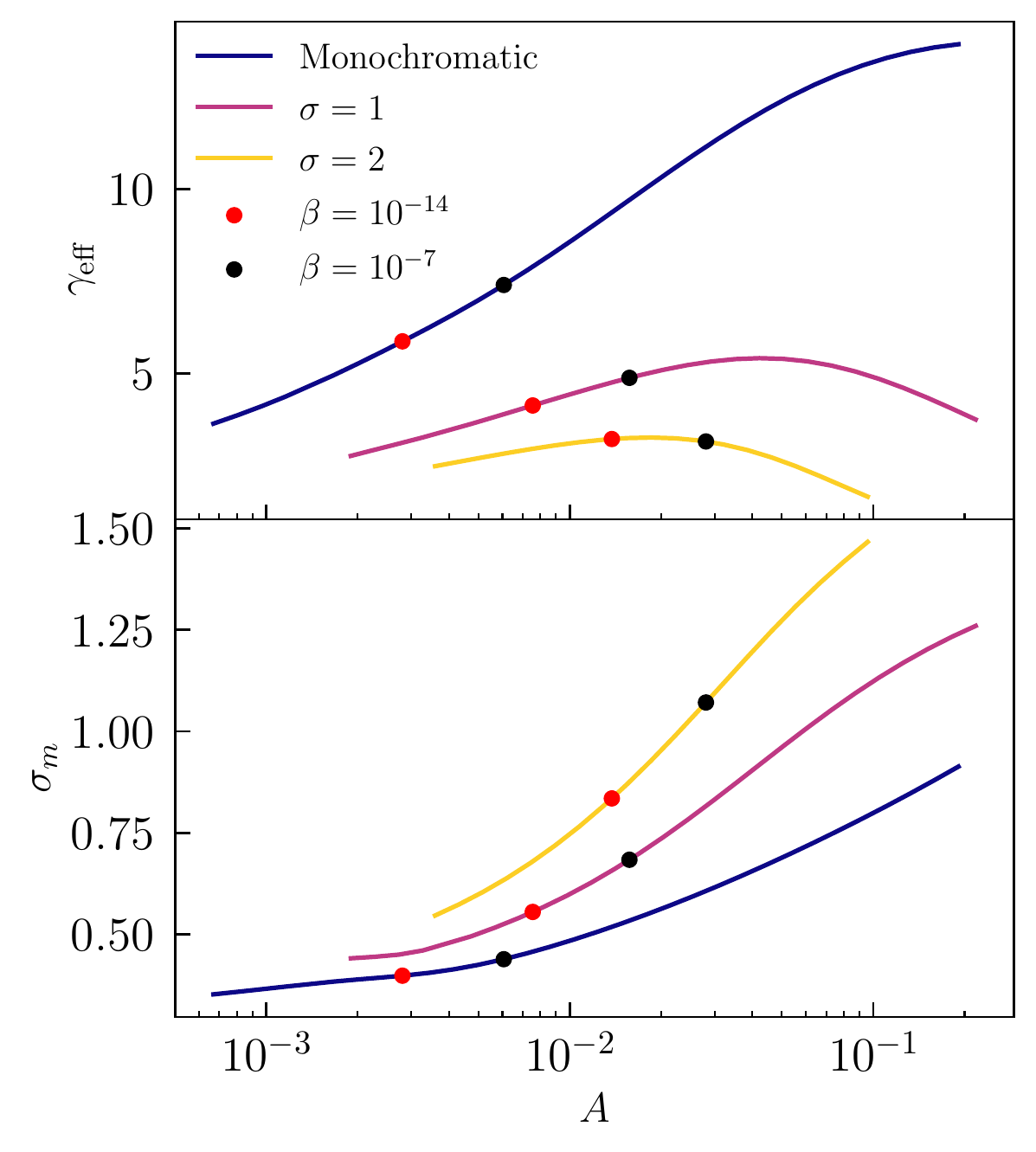}
    \caption{PBH mass function parameters as a function of the amplitude $A$ and width of the primordial curvature spectrum. The peak BH mass is $\gamma_{\rm eff} M_\star$ and the width of the log-normal distribution is $\sigma_m$. The dots indicate the $A$ values to produce $\beta_\star$ values of $10^{-7}$ and $10^{-14}$.}
    \label{fig:massfunction_dependence}
\end{figure}

Fig.~\ref{fig:massfunction_dependence} shows that $\gamma_\mathrm{eff}$ is an $O({\rm few})$ number that decreases with increasing curvature perturbation width and has a mild dependence on $A$.
The width of the mass function, $\sigma_m$, also grows with the curvature perturbation width and $A$. For example, in Fig.~\ref{fig:massfunction_dependence}, $\sigma_m$ ranges from 0.4 to 1.1 going from a monochromatic to $\sigma = 2$ primordial spectrum for typical $\beta$ values of interest for PBH observational constraints.
Studies of PBH constraints with log-normal mass functions have shown that for $\sigma_m \lesssim 0.2$, observational bounds on the PBHs are nearly identical to that of a monochromatic mass function~\cite{Carr:2017jsz}. However, the shape of the constraints starts to deviate from the monochromatic case for $\sigma_m \sim 1$ and will be substantially different as $\sigma_m$ is increased beyond 1. For this reason, we restrict to $\sigma < 2$ in the primordial power spectrum. Given the typical $\sigma_m$ values for these primordial perturbations, the observational bounds on monochromatic black hole mass functions that we show from \cite{Carr:2017jsz} are not exactly correct. In addition, some of the formulae in Sec.~\ref{sec:BHcosmo}, such as the early black hole domination condition, are also not exact in the presence of an extended mass function. We neglect these model-dependent $O(1)$ differences given that we are working with many orders of magnitude in the total parameter space.

%%%%%%%%%%%%%%%%%%%%%%%%%%%%%%%%%%%%%%%%%%
\section{Stochastic Gravitational Wave Spectrum}
\label{sec:SGWB}

There are potentially multiple sources of stochastic gravitational waves associated with these PBHs. Assuming PBHs form due to a large primordial scalar perturbation, then gravitational waves can be generated at second-order in perturbation theory. These gravitational waves are dominantly produced around the time of PBH formation, and will determine the sensitivity in the absence of a BHD era.  For sufficiently large density of PBHs such that there is a BHD era, then there are several additional effects. First, there can again be large scalar perturbations during the matter dominated-era due to the Poissonian distribution of PBHs, leading again to second-order gravitational waves. There may also be contributions from PBH clustering and evaporation at the end of a BHD era.  These sources from the BHD era do not depend on the origin of the PBHs being from scalar perturbations, but they do in principle depend on the resulting mass function of the PBHs. 

In this section, we detail each of these possible GW sources in turn. Note that aside from what is discussed here, there could also be GWs from Hawking radiation and PBH mergers~\cite{Raidal:2017mfl,Miller:2020kmv,DeLuca:2021hde,Zagorac:2019ekv,Inomata:2020lmk,Pujolas:2021yaw,Mukherjee:2021itf}, but these are generally at higher frequency and cannot be detected by GW interferometers for most of the PBH mass range we focus on. The GWs from mergers start to become detectable for some experiments at large masses near the very edge of our plots, but this contribution is more relevant for studies on solar mass PBHs.

\subsection{Second-order gravitational waves from primordial scalar perturbations}
\label{sec:GWs_from_primordial}

Primordial scalar perturbations provide a source for the tensor modes at second order in perturbation theory, as first noted in \cite{1967PThPh..37..831T}. In this section, we review the result following Ref.~\cite{Kohri:2018awv}, and also account for differences in the parameter space where there is a black hole dominated era. Again, we assume the primordial perturbations obey Gaussian statistics throughout. For discussion of GWs resulting from non-Gaussian primordial perturbations, see Refs.~\cite{Garcia-Bellido:2017aan,Cai:2018dig}.

As derived in detail in, \emph{e.g.,} Refs.~\cite{Baumann:2007zm,Kohri:2018awv,Espinosa:2018eve, Bugaev_2010,Saito_2010,Domenech:2019quo},  the equation of motion for each tensor mode polarization is given by
\begin{equation}
h_{\bfk}''(\eta) + 2 \mathcal{H} h_{\bfk}'(\eta) + k^2 h_{\bfk}(\eta) = 4 S_{\bfk}(\eta)
\label{eq:EOM}
\end{equation}
where $\eta$ is the conformal time, $\mathcal{H} = aH$ is conformal Hubble parameter, and $S_{\bfk}(\eta)$ is a source term which depends quadratically on the first order scalar perturbations $\Phi$. 
The tensor power spectrum and primordial power are related to the tensors and scalars by
\begin{align}
\mathcal{P}_\mathrm{GW} (\eta, k) \delta^3 (\textbf{k} + \textbf{k}') &= \frac{k^3}{2\pi^2} \langle h_\textbf{k} (\eta) h_{\textbf{k}'} (\eta) \rangle, \\
    \label{eq:power_spectrum_def}
\mathcal{P} (\eta, k) \delta^3 (\textbf{k} + \textbf{k}') &= \frac{k^3}{2\pi^2} \langle \Phi_\textbf{k} (\eta) \Phi_{\textbf{k}'} (\eta) \rangle.
\end{align}
The measurable quantity is the GW energy density fraction per logarithmic wavelength, which is given by
\begin{equation}
\Omega_{\mathrm{GW}} = \frac{1}{24}\left(\frac{k}{a(\eta)H(\eta)}\right)^2 \mathcal{P}_\mathrm{GW}(\eta,k)
    \label{eq:omega_gw_def}
\end{equation}
where we have summed over both polarization modes.

We begin by considering a long-lasting radiation dominated (RD) era after horizon entry of the primordial perturbation. During radiation domination, the source term drops as $1/a^\gamma$  with $\gamma \approx 3$ \cite{Baumann:2007zm} while for a freely propagating tensor mode $h \propto 1/a$. As a result, the power in the tensor modes will mainly be generated at horizon entry of the primordial perturbation, and well afterwards, they will behave as freely propagating gravitational waves. In this calculation, we will focus on those modes that enter the horizon during radiation domination and first study their evolution through RD. We will separately consider modes that enter the horizon during the PBH-dominated era below.

The GW dimensionless power spectrum in pure radiation-dominated (RD) is given by~\cite{Kohri:2018awv,Espinosa:2018eve,Domenech:2019quo}
\begin{align}
    \mathcal{P}_\mathrm{GW} (\eta,k) = 2 \int_0^\infty \!\!\!  dt & \int_{-1}^1 ds  \,  \mathcal{P}(kv) \mathcal{P}(ku) \, I^2 (s,t,x)\nonumber \\
    &\times \left[ \frac{t(2+t)(s^2 -1)}{(1-s+t)(1+s+t)}\right]^2
    \label{eq:RD_power_spectrum}
\end{align}
where $\mathcal{P}$ is the initial perturbation, $u$ and $v$ are defined as $u = \frac{t+s+1}{2}$ and $v = \frac{t-s+1}{2}$, and $x = \eta k$. The $I^2$ term arises from second order perturbation theory as an integral over some combination of the Green's function for tensors and the transfer functions for the scalar modes. To consider $k$ modes within the horizon such that the gravitational wave energy density is well-defined, we take the limit $x \to \infty$ (or $\eta k \gg 1$). In this limit and in a radiation-dominated era, $I^2$ can be explicitly written as 
\begin{align}
     &I^2(s,t,x)  =  \frac{288}{x^2} \frac{(s^2 + t(2+t) - 5)^4}{(s^2-(t + 1)^2)^6}\Bigg[\frac{\pi^2}{4} \Theta ( t - (\sqrt{3} - 1)) \nonumber \\
      & + \left( \frac{s^2-(t + 1)^2}{s^2 + t(2+t) - 5}+ \frac{1}{2} \log \left | \frac{-2+t(2+t)}{3-s^2} \right| \right)^2 \Bigg].
   \label{eq:I^2}
\end{align}
From this result, we see that the power spectrum  $\mathcal{P}_\mathrm{GW}$ scales as $\propto \frac{1}{\eta^2}$ in the $\eta k \gg 1$ limit. Since $a(\eta)H(\eta) = \frac{1}{\eta}$, we find that the GW energy density fraction in this limit is constant during RD:
\begin{equation}
    \Omega_\mathrm{GW}^{RD} = \frac{1}{24} (k \eta)^2 \mathcal{P}_{\rm GW}(\eta, k)|_{x \to \infty}
    \label{eq:omega_RD}
\end{equation}
corresponding to freely-propagating modes. Note that all dependence on the degrees of freedom is suppressed in the above equations and restored in the final expression below.

The GW spectrum in \eqref{eq:omega_RD} applies during the early radiation dominated era. If there is a BHD era before standard MRE, we must include an additional factor of $a(\eta_{\mathrm{BHD}})/a(\eta_{\mathrm{RH}})$ to account for the different scaling of the energy density in matter-dominated era. This factor is just the ratio of scale factors between the beginning of black hole domination and the period of reheating at the end of black hole domination.

Redshifting the energy density to today, we obtain the final result for the second-order GWs associated with the primordial perturbation: 
\begin{align}
      \Omega_\mathrm{GW} =  1.4\, \Omega_{\gamma,0}&
      \left(\frac{10.75}{g_\star}\right)^{1/3}  \frac{1}{24} (k \eta)^2 \mathcal{P}_{\rm GW}(\eta, k)|_{x \to \infty}
       \nonumber \\
      & \hspace{1cm} \times  \begin{cases}
      1  & \beta' < \beta'_{\rm min} \\
       \frac{a(\eta_{\mathrm{BHD}})}{a(\eta_{\mathrm{RH}}) } & \beta' \ge \beta'_{\rm min}
      \end{cases} .
    \label{eq:omegaGW_PBHformation}
\end{align}
$g_\star$ is degrees of freedom at the horizon entry of the $k_\star$ mode, and $\Omega_{\gamma, 0}$ is the abundance of photons today.
We emphasize again that this applies for modes which have entered the horizon before PBH domination and therefore satisfy $k \eta_{\mathrm{BHD}} \gtrsim 1$, or equivalently $k \gtrsim \beta k_\star$. 
For the GWs produced at PBH formation, the experimental reach is primarily from tensor modes that are near the peak mode $k_\star$ or a few orders of magnitude below, so this result will suffice for our purposes.

%%%%%%%%%%%%%%%%%%%%%%%%%%%%%%%%%%%%%%%%%%

\begin{figure}
    \includegraphics[width=0.98\linewidth]{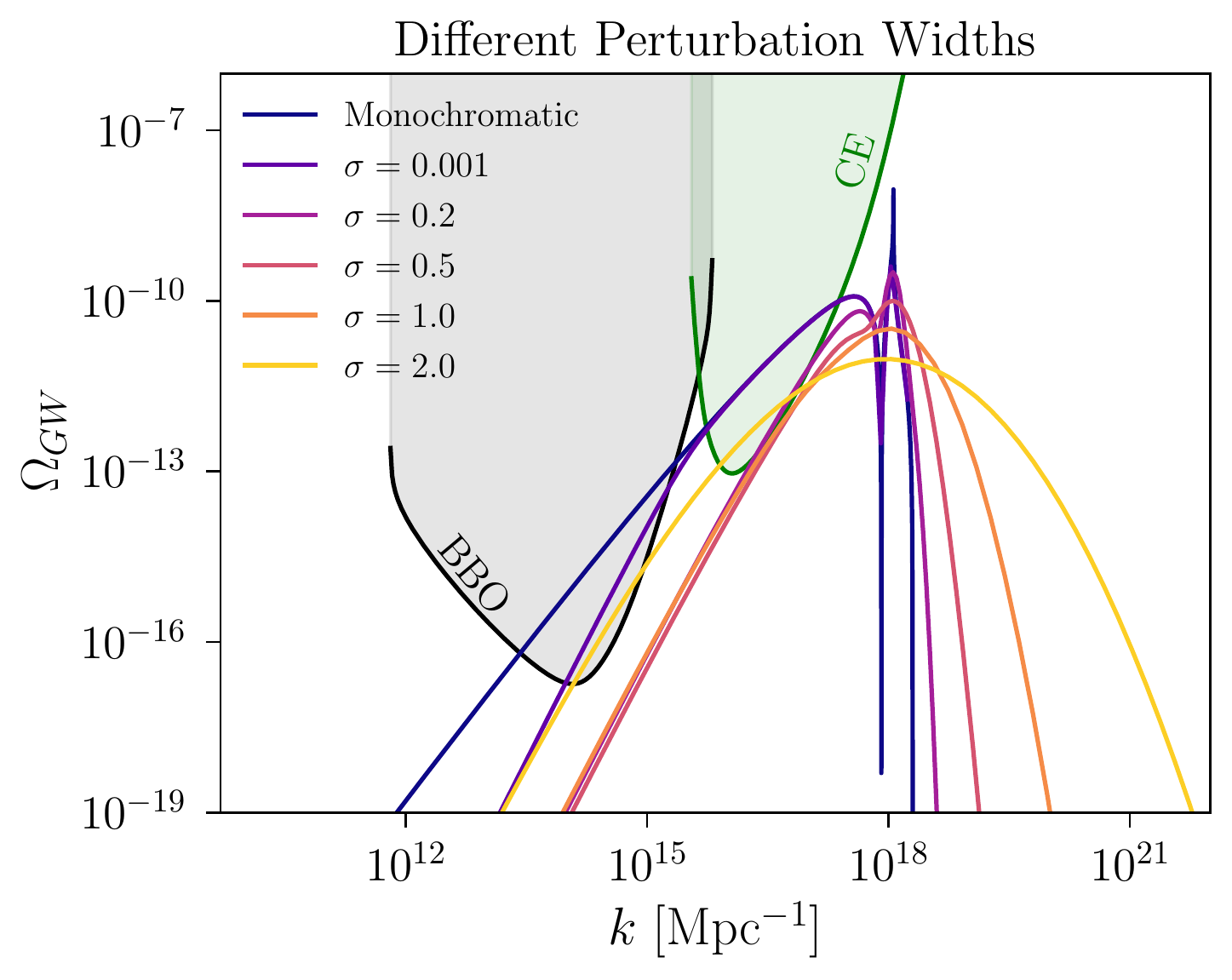}
    \caption{Gravitational wave spectrum induced at second order from the primordial scalar perturbations. $\Omega_{\rm GW}$ is the energy density per logarithmic wavelength today. The different lines show the dependence on the width of primordial curvature power spectrum, $\sigma$. For reference, we take $A = 0.003$, $k_\star = 10^{18}~\Mpc^{-1}$. The black and green lines are power law integrated sensitivity curves for BBO and CE; for more details see Sec.~\ref{sec:reach}.   }
    \label{fig:different_sigma}
\end{figure}

%%%%%%%%%%%%%%%%%%%%%%%%%%%%%%%%%%%%%%%%%%

%%%%%%%%%%%%%%%%%%%%%%%%%%%%%%%%%%%%%%%%%%
\subsubsection{Dependence on primordial perturbation}

\label{subsec:primordial_perturbation}

For sufficiently narrow primordial spectra, the BH mass function will not depend much on the details of the shape or width of the spectrum. The situation is quite different for the GW spectrum discussed in this section, however, where the infrared tail is quite sensitive to the width of the primordial spectrum.

Fig.~\ref{fig:different_sigma} shows the GW spectrum for a monochromatic perturbation as well as Gaussians with a range of widths. For the monochromatic case, the spectrum goes as $k^2$ for $k \ll k_\star$. However, for a Gaussian primordial perturbation with width $\sigma$, the slope of the GW spectrum goes as $k^2$ for $\sigma k_\star \lesssim k \lesssim k_\star$, and goes as $k^3$ for $k \lesssim \sigma k_\star$ as also shown in \cite{PhysRevD.101.043019}. The $k^3$ tail is a generic feature of gravitational waves produced at second order in a radiation dominated universe, as long as we have perturbations satisfying the conditions given in Ref.~\cite{Cai:2019cdl}. The important conditions here are that the source has finite width and that the $k$ must be less than all characteristic scales associated with the source. The monochromatic case of course never satisfies these conditions, so it does not show this $k^3$ infrared scaling. In the finite width case, the condition that $k$ is less than the characteristic scales associated with the source is satisfied for $k < \sigma k_\star$. This can be seen in the change in $k$ scaling at $k \sim \sigma k_\star$ for the case $\sigma = 0.001$ in Fig.~\ref{fig:different_sigma}. For the other $O(1)$ values of $\sigma$, the possible region with $k^2$ scaling is not visible and instead the dominant behavior of the tail is the $k^3$ scaling. The width-dependence of the GW spectrum can thus have large impacts on GW detectability, even when the BH mass functions are similar.

In the monochromatic case, we can see the behavior for $k \ll k_\star$ by simply evaluating the integral \eqref{eq:RD_power_spectrum} at the peak, where $s=0$, $t=2k/k_\star - 1$. Defining $\tilde{k}=k/k_\star$, then for $\tilde k \ll 1$, we have 
\begin{equation}
 \Omega^{RD}_\mathrm{GW} = A^2 \frac{3}{4} \tilde{k}^2 \left[\log ^2 \! \left(\frac{4}{3 \tilde{k}^2}\right)-4 \log \!
   \left(\frac{4}{3 \tilde{k}^2}\right)+\pi ^2+4\right]. \nonumber
\end{equation}
In the Gaussian case, we can analytically approximate the result via the Laplace method. Expanding the integrand about the peak $s=0$, $t=\frac{2k_\star}{k} e^{-2\sigma^2} - 1$ and evaluating gives the following simple expression for the power spectrum for $\tilde k < \sigma$:
\begin{align}
 \Omega^{RD}_\mathrm{GW}& = A^2 \frac{3 {\tilde{k}}^3 e^{2 \sigma ^2} \log ^3 \! \left(\frac{4 e^{-4 \sigma ^2}}{3 {\tilde{k}}^2}\right) }{8 \sqrt{2} \sigma } \times \\
 &\left[
 \log^2 \! \left(\frac{4 e^{-4 \sigma ^2}}{3 {\tilde{k}}^2}\right)+2 \sigma ^2 \log \! \left(\frac{4
   e^{-4 \sigma ^2}}{3 {\tilde{k}}^2}\right)+4 \sigma ^2 \right]^{-1/2} \nonumber
\end{align}
which indeed goes as $\tilde k^3$.

%%%%%%%%%%%%%%%%%%%%%%%%%%%%%%%%%%%%%%%%%%
\subsection{Gravitational waves from PBH-dominated era}

\label{sec:GWs_from_BHD}

In addition to GWs produced at PBH formation,  which dominates near the formation scale $k_\star$, there are also GWs generated during matter domination~\cite{Papanikolaou:2020} that contribute at $k$ below $k_\star$. 
We first consider GWs that arise from second  order perturbation theory sourced by the scalar perturbations, with the difference in this case that the scalar perturbations are produced by the gravitational potential of the PBHs themselves.

We will adapt our estimate from \cite{Papanikolaou:2020}, including here an additional correction factor to restrict to the linear regime in perturbation theory. We first review Ref.~\cite{Papanikolaou:2020}, which  directly computes the curvature power spectrum assuming a Poissonian distributed gas of PBHs. At formation, the density contrast in black holes can be treated as an isocurvature perturbation, which later evolves into a curvature perturbation. From \cite{Papanikolaou:2020}, we have the following approximation for the perturbation at the onset of BHD:
\begin{equation}
    \mathcal{P}_\Phi (k) = \frac{2}{3 \pi} \left (\frac{k}{k_\mathrm{UV}}\right)^3 \left(5 + \frac{4}{9} \frac{k^2}{\mathcal{H}_{\mathrm{d}}^2} \right)^{-2}
    \label{eq:curvature_power_BHD_era}
\end{equation}
where $\mathcal{H}_\mathrm{d}$ is comoving Hubble at the onset of BHD and $k_\mathrm{UV}$ is the mean comoving BH separation scale,
\begin{equation}
    k_\mathrm{UV} = \left(\frac{\beta}{\gamma_\mathrm{eff}}\right)^{1/3} k_\star.
    \label{eq:kUV}
\end{equation}
Upon PBH formation, the black hole energy fraction grows, and the comoving Hubble when the black hole dominated era begins is $\mathcal{H}_\mathrm{d}=\beta k_\star$.

In order to calculate the contribution to the tensor power spectrum, we  make the replacements $\mathcal{P} \rightarrow \mathcal{P}_\Phi$ and $I \rightarrow I_\mathrm{MD}$ in (\ref{eq:RD_power_spectrum}).
Recall $I$ arose in second order perturbation theory as an integral over a combination of the Green's function in RD and quadratic terms in the transfer function, so we must replace these with the solutions for the PBH era. This is the same as an matter-dominated (MD) era for our GW calculation, since BHs can be treated as a pressure-less non-relativistic fluid for modes $ < k_{\rm UV} \ll k_\star$. The most important difference to note between the RD and MD eras is that the transfer function decays in RD but is constant in MD, so in the limit $x \rightarrow \infty$, $I^2_\mathrm{RD} \propto \frac{1}{x^2}$ as in (\ref{eq:I^2}), while $I^2_\mathrm{MD} = \mathrm{const}$. The source is constant and amplifies the resulting GWs from the BHD era.

For the BHD era, the energy density is $\Omega_{\rm GW} = (k/a H)^2 \mathcal{P}_{\rm GW}(k) / 48$, which has a factor of 2 difference from the corresponding relationship for GWs generated in the RD era, \eqref{eq:omega_gw_def}. While in the RD era there is an equal contribution from kinetic and gradient terms, in the MD era the kinetic terms are negligible when there is a constant source term. This is because the constant source forces the tensor modes to constant values at late times in the MD era. An approximate form for the GW spectrum today is then given by~\cite{Papanikolaou:2020}
\begin{align}
    \Omega_\mathrm{GW} = &  \,  4.4 \times 10^{19}\, \nonumber \Omega_{\gamma,0} \times  \left(\frac{g_\mathrm{BHD}}{106.75}\right)^{-1/3}   \left(\frac{g_\mathrm{RH}}{10.75}\right)^{-1/3} \nonumber \\
     &\times 
    \left(\frac{m_\mathrm{BH}}{10^9 \ \mathrm{g}} \right)^{4/3} (\beta')^{16/3} \nonumber \\
    & \times  \begin{cases}
    \frac{k}{\mathcal{H}_\mathrm{d}} & k < 8\mathcal{H}_\mathrm{d} \\
    8 & 8\mathcal{H}_\mathrm{d} < k < k_\mathrm{UV} \\
    0 & k > k_\mathrm{UV}
    \end{cases}
    .
\label{eq:omega_bhd}
\end{align}
Here we have assumed that the gravitational waves redshift as radiation after the transition from the PBH-dominated era to the radiation era. In general, there can be additional dampening or growth in the GW spectrum result depending on the details of the transition between the PBH and radiation dominated eras~\cite{Inomata:2019zqy,PhysRevD.100.043532}. For example, if there is a sufficiently narrow BH mass function, there is a rapid evaporation and a sudden change in the equation of state at the end of the MD era, leading to a significant enhancement in the spectrum relative to \eqref{eq:omega_bhd}~\cite{PhysRevD.100.043532, Domenech:2020,Domenech:2021wkk}. This is the case if the BH mass function has $\sigma_m \lesssim 0.01$ \cite{Inomata:2020lmk}, while we have log-normal mass functions with typical $\sigma_m \sim 0.4-1.2$.

However, it is not clear whether \eqref{eq:omega_bhd} applies for an extended MD era, because at some point during the MD era, density perturbations can become nonlinear. The density perturbation at black hole formation is given by
\begin{equation}
    \delta(t_\mathrm{BHD}) \sim \sqrt{\frac{2}{3 \pi}} \left(\frac{k}{k_\mathrm{UV}}\right)^{3/2}
    \label{eq:delta_formation}
\end{equation}
and starts growing with the scale factor at $\mathrm{Max}[a_\mathrm{BHD}, a_k]$, where $a_k$ is the scale factor at horizon entry for that mode, since the density perturbation is frozen while outside of the horizon and can only grow in the BHD era. Once perturbations become nonlinear, the perturbation theory solution of the scalar modes with a constant source term is no longer valid. There may be GWs produced instead by collapse of nonlinear perturbations or mergers, as estimated in \cite{Jedamzik:2010hq}, but these require further numerical simulations to treat properly. 

A lower bound on the possible GW spectrum can be obtained by restricting the source function to regions in $k$ and time for which $\delta_k(t) < 1$. 
Setting the cutoff at $\delta_k = 1$ gives the cutoff scale factor $a_\mathrm{cut}(k)$:
\begin{equation}
\frac{a_\mathrm{cut}}{a_\mathrm{RH}} = \frac{a_\mathrm{BHD}}{a_\mathrm{RH}} \times
\begin{cases}
 \sqrt{\frac{3\pi}{2}} \left( \frac{\mathcal{H}_\mathrm{d}}{k}\right)^2\left(\frac{k_\mathrm{UV}}{k}\right)^{3/2} & k < \mathcal{H}_\mathrm{d} \\
 \sqrt{\frac{3\pi}{2}} \left(\frac{k_\mathrm{UV}}{k}\right)^{3/2} & k > \mathcal{H}_\mathrm{d} \\
\end{cases}.
\label{eq:nonlinear_cut_redshift}
\end{equation}
Compared to the results of \cite{Papanikolaou:2020}, cutting off the source function when density perturbations become nonlinear introduces an additional factor in \eqref{eq:omega_bhd} given by
\begin{equation}
R \approx \Bigg( \mathrm{Min} \Bigg[ \frac{a_\mathrm{BHD}}{a_\mathrm{RH}} \times
\begin{cases}
 \sqrt{\frac{3\pi}{2}} \left(\frac{k_\mathrm{UV}}{6 k_\mathrm{BHD}}\right)^{3/2} & k < 6 \mathcal{H}_\mathrm{d} \\
 \sqrt{\frac{3\pi}{2}} \left(\frac{k_\mathrm{UV}}{k}\right)^{3/2} & k > 6 \mathcal{H}_\mathrm{d} \\
\end{cases}, 1 \Bigg] \Bigg)^2, \label{eq:bhd_Rfactor}
\end{equation}
see Appendix \ref{sec:nonlinear_cutoff} for details.
Here we have connected the results at $k \ll \mathcal{H}_\mathrm{d}$ and $k \gg \mathcal{H}_\mathrm{d}$ as the details in the intermediate regime are unimportant. With this approach, the source function is cut off well before the transition to radiation domination, leading to a strong dampening of the spectrum. Then the spectrum is far beneath the threshold required for experimental observability, such the details of how the source is cut off and the transition from the PBH to the RD era are not relevant for our calculation. We note that our estimate may be overly conservative, however.
In addition, other sources of GWs will be present and can be many orders of magnitude larger when the perturbations are nonlinear~\cite{Jedamzik:2010hq}. We estimate one possible source of GWs generated at the end of the BHD era in the following subsection. 

\subsection{Gravitational waves from reheating \label{sec:GWs_from_reheat}}

There can further be contributions to GWs from reheating of the universe due to PBH evaporation. One possibility is that GWs are produced when the nonlinear structures of the BHD era all evaporate into radiation. The idea is that the black holes slowly evaporate a fraction of their mass into  radiation through the bulk of the BHD era. Then, at the end of the BHD era, the remaining mass contained within the halos evaporates and collides against the radiation fluid.  This could produce a turbulent fluid with eddies that cascade down to smaller scales, which can induce GWs. In this section, we perform an order of magnitude estimate for this potential source. Detailed numerical simulations are required to determine if evaporation of the clustered PBHs indeed acts as an efficient source of turbulence. 

The topic of GWs from turbulent events in the early universe has been studied in Refs.~\cite{Kosowsky:2001xp, Gogoberidze:2007an}. Here we discuss the assumptions and calculations in these works and argue that the results can be used in estimating the GWs sourced by the PBHs at the end of their lifetime. In these studies, the picture is that during the early universe, a fraction of the available energy density is transformed into kinetic energy of the cosmological fluid. This kinetic energy is stirred on a length scale $L_S$ over a time scale $\tau_\mathrm{stir}$. These quantities correspond to the characteristic length and duration of the turbulent source. The turbulent kinetic energy then cascades down to some damping scale determined by the fluid characteristics of the plasma.

In these works, the turbulent source is left generic and the only assumption is that the spectrum of the turbulence is Kolmogorov. We note that the Kolmogorov energy spectrum is a result from classical hydrodynamics and that relativistic turbulence is much less explored; nonetheless, we follow other works in assuming that classical theory provides an approximation for turbulence in a relativistic fluid.
To summarize, the calculations of these studies depend on the scale and duration of turbulence, cosmological parameters and characteristics of the plasma during energy injection, and the model of the turbulence. The gravitational wave spectrum from turbulence is then calculated by solving the usual wave equation for  tensor modes sourced by the transverse traceless piece of the stress-energy tensor. Refs.~\cite{Kosowsky:2001xp} and \cite{Gogoberidze:2007an} perform this analysis in Fourier and real space, respectively, arriving at approximately the same results with slight differences arising from different treatments in the time dependence of the turbulence. In particular, no assumption is made on the particle physics origin of the turbulence. For example, while Refs.~\cite{Kosowsky:2001xp, Gogoberidze:2007an} discussed phase transitions as one possible source of GWs, the results have been applied to other scenarios such as decay of a scalar field reheating the universe~\cite{Jedamzik:2010hq}.

We may apply these results to calculate GWs from Hawking evaporation in the BHD era, given the following physical picture. For the bulk of the BHD era, the universe will look like clusters of black holes with sparse radiation fluid freely streaming outwards. However, by the end of the BHD era, all of the mass in the black hole clusters then gets deposited into radiation, which is a possible source of turbulence. Since we expect the typical black hole mass function to have a fairly significant width as in Fig.~\ref{fig:mass_function}, we approximate the turbulent source to last for around $\tau_\mathrm{stir} \approx t_{\rm evap}$. 
In addition, we take the length scale of the turbulence source $L_S$ to be defined by the smallest comoving wavenumber that becomes non-linear by black hole evaporation, $k_\mathrm{NL}$. That is, $k_\mathrm{NL}$ is the comoving wavenumber that satisfies $a_\mathrm{cut} = a_\mathrm{RH}$ in (\ref{eq:nonlinear_cut_redshift}), which gives
\begin{equation}
k_\mathrm{NL} = 
\begin{cases}
 \frac{3.2 \times 10^{14}}{\mathrm{Mpc}} \big( \frac{\MBH}{10^4 \mathrm{g} } \big)^{-\frac{17}{14}}  & \beta' > \beta'_\mathrm{NL} \\
 \frac{1 \times 10^{16}}{\mathrm{Mpc}} \big( \frac{\MBH}{10^4 \mathrm{g} } \big)^{-\frac{31}{18}} \big( \frac{\beta'}{10^{-7}} \big)^{-\frac{8}{9}} & \beta' < \beta'_\mathrm{NL} \\
\end{cases}
\label{eq:k_NL}
\end{equation}
where
\begin{equation}
\beta'_\mathrm{NL} = 5 \times 10^{-6} \left(\frac{\MBH}{10^4 \mathrm{g}} \right)^{-\frac{4}{7}}
\end{equation}
is the minimum $\beta'$ such that the largest non-linear mode enters the horizon after the BHD era starts.

As argued in \cite{Kosowsky:2001xp}, the relevant time-scale for production of gravitational waves from turbulence is given by the maximum of the turbulence duration $\tau_\mathrm{stir}$ and the dissipation time of the largest eddies $\tau_S$. In the PBH evaporation scenario, the black hole mass function is wide enough that the larger time-scale is on the order of the black hole evaporation time $t_\mathrm{evap}$. Then, the energy dissipation rate is given by 
\begin{equation}
\label{eq:energy_dissipation}
    \epsilon \sim \frac{\kappa \rho}{w t_{\rm evap}},
\end{equation}
where $\rho$ is the energy density, $\kappa$ is the efficiency factor of conversion to turbulent kinetic energy, and $w$ is the enthalpy density of the radiation fluid. Note that this is highly approximate, as we've effectively treated the turbulence as uniformly injected over the time $t_{\rm evap}$ rather than modeling the exact time-dependence of the Hawking radiation, so the result should be treated as a very rough estimate. Another important quantity for the turbulence calculation is the Mach number defined in Ref.~\cite{Gogoberidze:2007an}, which goes as 
\begin{equation}
    M \propto \Big( \frac{\epsilon}{k_{\mathrm{NL}}} \Big)^{1/3} \propto \kappa^{1/3} \Big( \frac{k_\mathrm{RH}}{k_\mathrm{NL}} \Big)^{1/3}.
\end{equation}
Typical Mach numbers for these early evaporating black holes are $M \sim 0.1 - 0.5$. The comoving wavenumber of the horizon at the end of evaporation $k_\mathrm{RH}$ can be calculated from (\ref{eq:t_evap}) and is given by
\begin{equation}
    \label{eq:k_RH}
    k_\mathrm{RH} = \frac{4.9 \times 10^{11}}{\Mpc} \Big(\frac{\MBH}{10^4} \Big)^{-\frac{3}{2}}.
\end{equation}

 With these physical quantities in hand, we may apply the final result of Ref.~\cite{Gogoberidze:2007an} to our scenario. For clarity, we rewrite the original expression here and make explicit the replacements in our variables. The turbulence-sourced GW spectrum today is~\cite{Gogoberidze:2007an}
\begin{align}
    \label{eq:omega_turbulence_ref}
    \nonumber
    h_c\propto& \Big(\frac{100\,  \GeV}{T_*} \Big) \Big(\frac{100}{g_*} \Big)^{1/3} \times \\ & \Big( \frac{H_*}{k_0} \Big)^{3/2} (H_* \tau_\mathrm{stir})^{1/2} \Big( k_0^3 \omega H_{ijij} (\omega, \omega) \Big)^{1/2},
\end{align}
where $T_*$, $g_*$ are the temperature and degrees of freedom at the end of the turbulence,  $k_0$ is the inverse length scale of the turbulent source, $\tau_\mathrm{stir}$ is the turbulence duration, $\omega$ is the angular frequency of the gravitational waves at $t_\mathrm{evap}$,
and the function $H_{ijij}$ is approximated by
\begin{equation}
\frac{7 M^3 k_0^{-4}}{16 \pi^{3/2}} \int_0^1 dx \, x^{11/4} \mathrm{exp} \Big[-\Big(\frac{\omega}{k_0 M}\Big)^2 x \Big] \mathrm{Erfc} \Big[-\frac{\omega}{k_0 M} \sqrt{x} \Big]. \nonumber
\end{equation}
Then, taking $T_* = T_\mathrm{RH}$, $g_* = g_\mathrm{RH}$, $\tau_{\mathrm{stir}} = t_\mathrm{evap}$, and rewriting in terms of comoving quantities $H_* = k_\mathrm{RH}/a_\mathrm{RH}$, $k_0 = k_\mathrm{NL}/a_\mathrm{RH}$, $\omega = k_\mathrm{NL}/a_\mathrm{RH}$, and using the usual  $\Omega_\mathrm{GW} = \frac{2\pi^2}{3H_)^2} f^2 h_c^2$, we have
\begin{align}
    \label{eq:omega_turbulence}
    \nonumber
    \Omega_\mathrm{GW,turb} =& 4 \times 10^{-22} \Big(\frac{100\,  \GeV}{\TRH} \Big)^2 \Big(\frac{100}{g_\mathrm{RH}} \Big)^{2/3} \times \\ & \Big( \frac{k}{1 \Mpc^{-1}} \Big)^2 \Big(\frac{k_\mathrm{RH}}{k_\mathrm{NL}} \Big)^4 \Big(\frac{k}{k_{\mathrm{NL}}} \times F(k) \Big )
\end{align}
where $F(k)$ is given by
\begin{equation}
\frac{7 M^3}{16 \pi^{3/2}} \int_0^1 dx \, x^{11/4} \mathrm{exp} \Big[-\Big(\frac{k}{k_\mathrm{NL} M}\Big)^2 x \Big] \mathrm{Erfc} \Big[-\frac{k}{k_\mathrm{NL} M} \sqrt{x} \Big]. \nonumber
\end{equation}
The function $F(k)$ is approximately constant below $k_\mathrm{NL} M$ and decays as $k^{-15/2}$ above $k_\mathrm{NL} M$, so the GW spectrum from turbulence peaks at $k_\mathrm{NL} M$. When the initial PBH abundance is well above the BHD line, the spectrum has no $\beta'$ dependence. This is because at large $\beta'$, the black hole dominated era starts early enough that $k_{\rm NL}$ is only dependent on the PBH mass. This can  be seen from \eqref{eq:k_NL}, which is $\beta'$ independent above $\beta'_{\rm NL}$. Within 3-4 orders of magnitude in $\beta'$ of the BHD line, there is a large suppression in this spectrum since the BHD era is much shorter and the halo sizes are smaller.

The spectrum in (\ref{eq:omega_turbulence}) should be regarded only as a rough estimate of the GWs from turbulence. For example, a large uncertainty arises from the dependence on the efficiency factor $\kappa$. There should also be an additional suppression from the expansion of the universe since the turbulence occurs over a Hubble time. Regardless, even with an optimistic efficiency factor $\kappa = 1$, our turbulence estimate in (\ref{eq:omega_turbulence}) is not strong enough to be detected even by BBO and does not appear in our reach plots.

%%%%%%%%%%%%%%%%%%%%%%%%%%%%%%%%%%%%%%%%%%

\begin{figure}[t]
    \centering
    \includegraphics[width=0.99\linewidth]{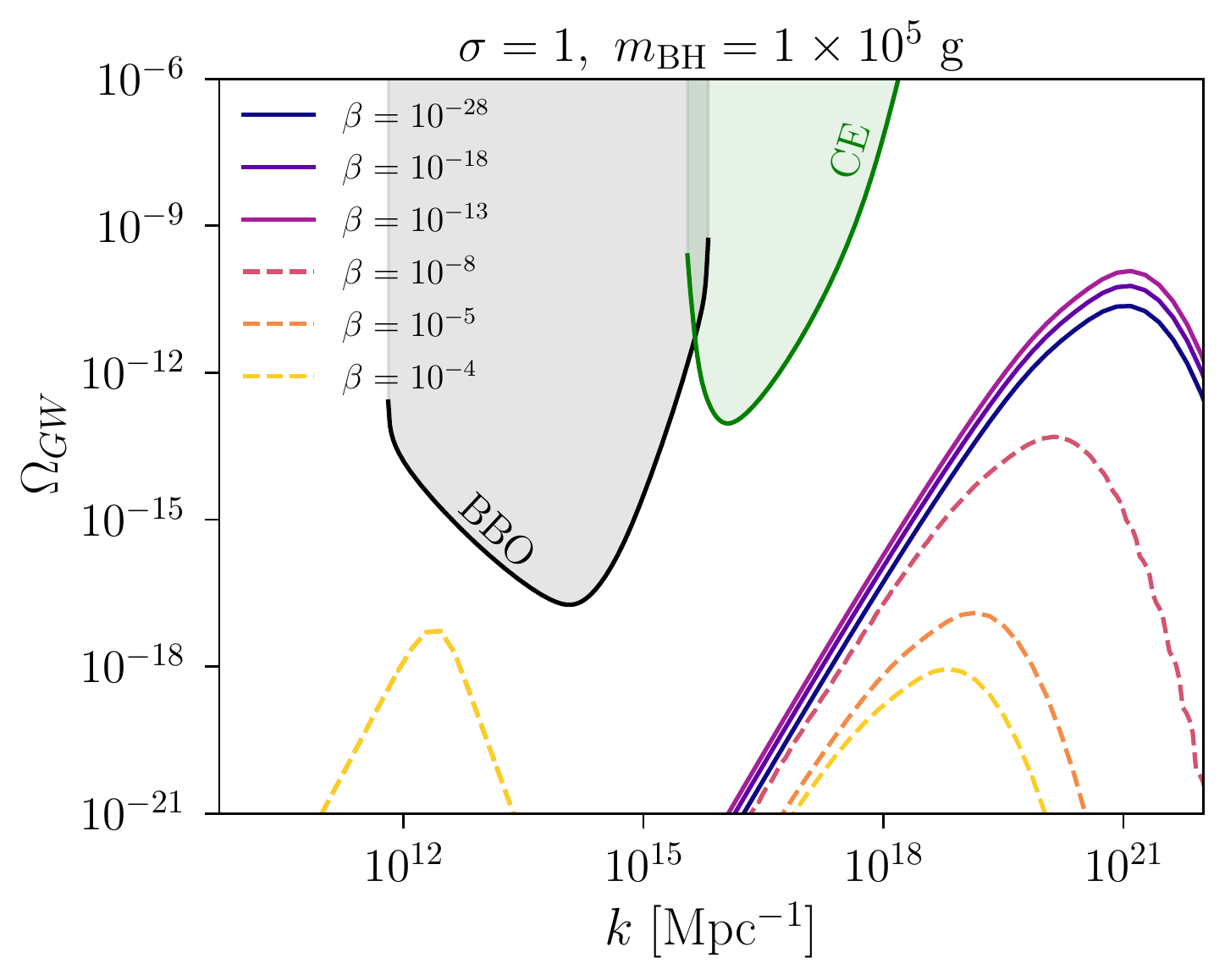}
    \caption{Total gravitational wave spectrum today from a Gaussian perturbation with $\sigma=1$, for $m_\mathrm{BH}=1 \times 10^5$g and different values of $\beta$. The solid lines are for small $\beta$ with no BHD era. The dashed lines indicate the total resulting GWs when there is a BHD era. We see that when $\beta$ increases, the dominant contribution shifts to the GWs from turbulence as discussed in Sec.~\ref{sec:GWs_from_reheat}. For this contribution, we assume an optimal efficiency of $\kappa = 1$, which gives a Mach number of $M \approx 0.2$. For $\beta$ much larger than the minimum required for black hole domination, the turbulent contribution is identical. 
    }
    \label{fig:OmegaGW_different_beta}
\end{figure}

%%%%%%%%%%%%%%%%%%%%%%%%%%%%%%%%%%%%%%%%%%

\subsection{Combined spectrum}

In Fig.~\ref{fig:OmegaGW_different_beta}, we show the combined GW spectrum for a reference PBH mass of $m_\mathrm{BH}=1 \times 10^5$g and assuming a primordial perturbation with lognormal width $\sigma=1$. For small enough $\beta$, there is no BHD era, and the only GWs are those induced from the primordial perturbation, discussed in Sec.~\ref{sec:GWs_from_primordial}. The dashed lines show the GW spectrum where there is a BHD era.  The BHD era leads to an overall redshifting of the GW spectrum associated with BH formation. For this signal, it also leads to a peak frequency $k_\star$ which depends on $\beta$ according to \eqref{eq:kstar_BHD}, leading to a peak at lower frequencies as $\beta$ is increased. In addition, our rough estimate of the GW spectrum generated by turbulence at PBH evaporation is visible, giving a contribution that peaks at lower frequencies. 
The contribution from gravitational waves generated during the BHD era discussed in Sec.~\ref{sec:GWs_from_BHD} peaks in a similar frequency range as the turbulent source, but has negligible amplitude when we implement a cutoff on nonlinear density perturbations. We emphasize again that our estimate of this latter contribution is likely too conservative and that a numerical study is required to calculate the GW spectrum from the black hole dominated non-linear regime.

%%%%%%%%%%%%%%%%%%%%%%%%%%%%%%%%%%%%%%%%%%
\section{Observational Bounds and Reach \label{sec:reach} }

In this section we discuss the experimental reach for various GW experiments on the stochastic GW background associated with PBHs. We show the sensitivity in terms of the parameter space for the primordial scalar perturbation, $(k_\star, A)$, as well as in the black hole mass and mass fraction parameter space,  $(m_\mathrm{BH}, \beta')$. We also translate existing PBH bounds into the $(k_\star,A)$ plane.

\begin{figure*}[t]
    \centering
    \includegraphics[width=0.47\linewidth]{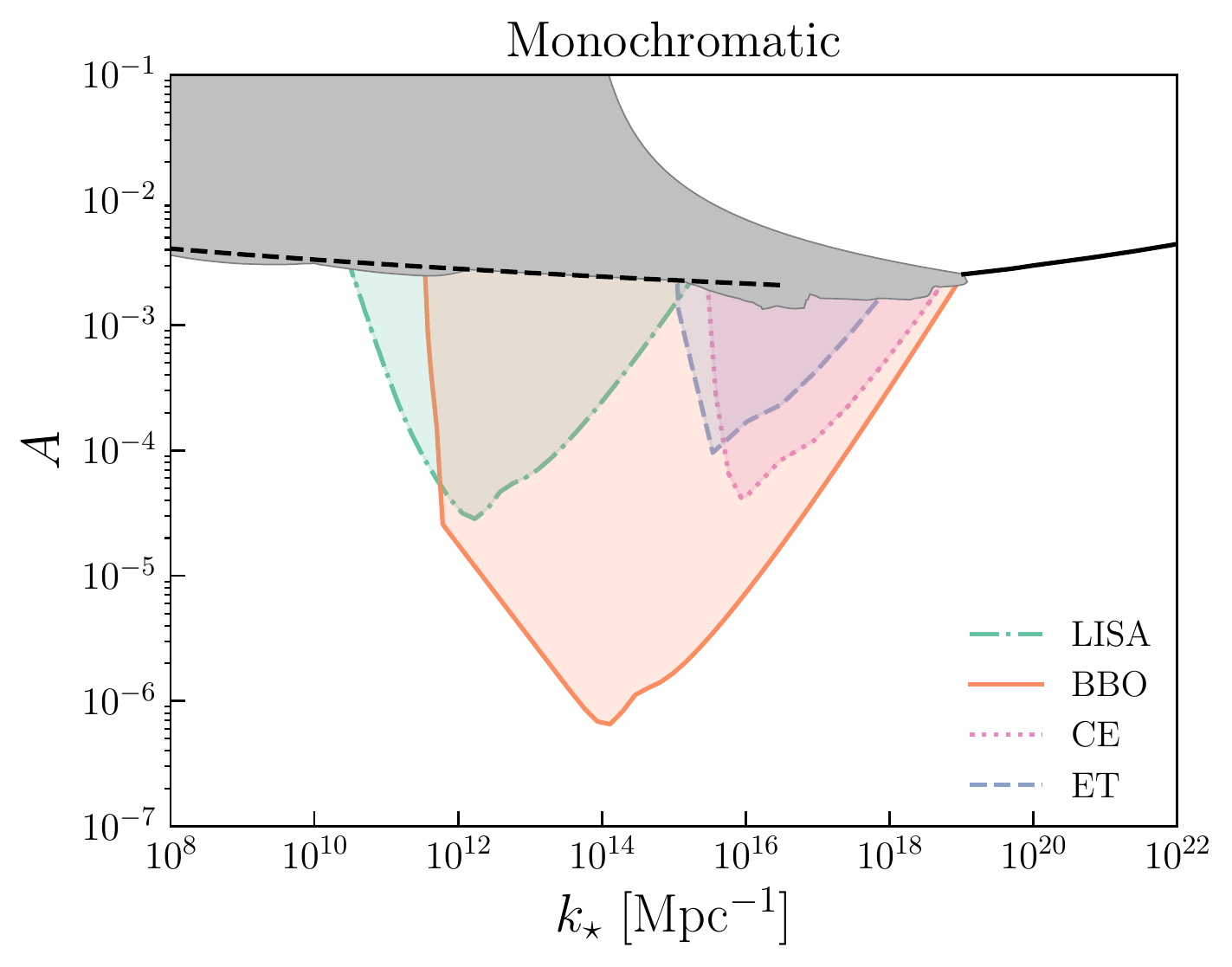}
    \includegraphics[width=0.47\linewidth]{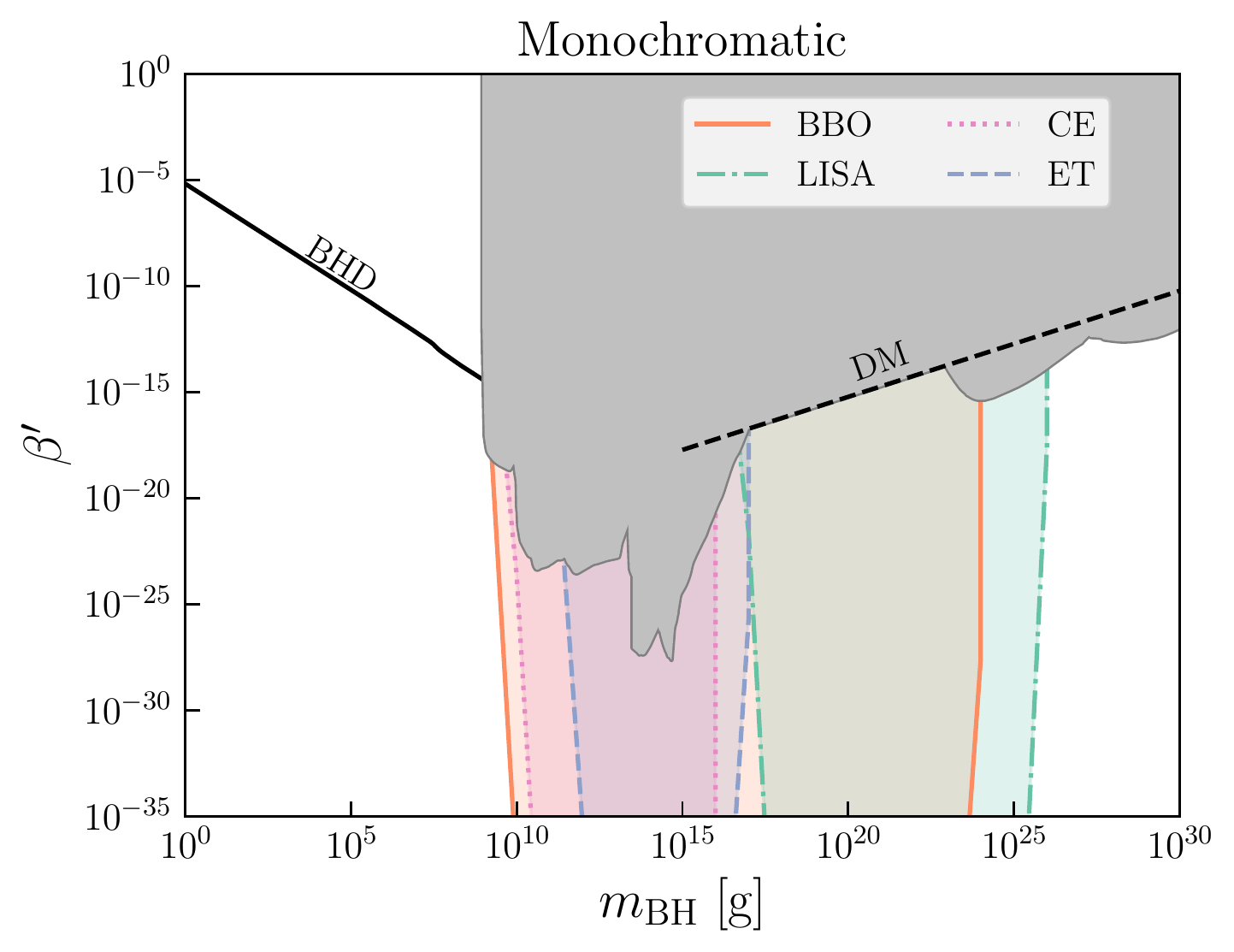}
\caption{For monochromatic perturbations, the observational reach for experiments BBO, LISA, CE, ET are shown in the $(k_\star, A)$ plane for the primordial scalar perturbation ({\bf left}) and in the $(\beta', m_{\rm BH})$ plane of PBH parameters ({\bf right}).  We assume an observational time of 1 year and a signal-to-noise ratio of 1 for each of the experiments. In both panels, the gray shaded regions are the existing PBH bounds that were shown in Fig.~\ref{fig:BHD_regions} and discussed in Sec.~\ref{sec:constraints}.  The DM line (black, dashed) shows the parameters that would result in the saturation of current observed dark matter density today. The BHD line (black, solid)  is the minimum $\beta'$ at a given $m_\mathrm{BH}$ that would result in a period of black hole domination. 
Above the BHD line, none of the GW sources we have estimated are observable.
%we estimate the experimental sensitivity to GWs produced by the turbulent fluid at the end of the BHD era.
For $m_\mathrm{BH}>10^9$ g, the experimental reach comes from the second-order GWs induced by the primordial scalar perturbation. Note that pulsar timing arrays (PTA) have some observational reach near at $k_\star \sim 10^8 \ \mathrm{Mpc}^{-1}$ or $m_\mathrm{BH} \sim 10^{30}$ g, but we choose not to show this here as these reaches are most relevant for near solar mass PBHs. 
}
    \label{fig:monochromatic_bounds}
\end{figure*}

\subsection{Calculation of Observational Bounds}

We calculate the expected observational bounds and observational reach from various current and proposed experiments following Refs.~\cite{Thrane:2013oya,Schmitz:2020syl}. We consider Big Bang Observer (BBO) \cite{Crowder:2005nr,Corbin:2005ny,Harry:2006fi,Yagi:2011wg}, Einstein Telescope (ET) \cite{Maggiore:2019uih,Punturo:2010zz}, Cosmic Explorer (CE) \cite{Reitze:2019iox,LIGOScientific:2016wof}, and the Laser Interferometer Space Antenna (LISA) \cite{Audley:2017drz,Baker:2019nia}. Other proposed interferometers that operate in a similar frequency range to LISA include TianQin~\cite{TianQin:2015yph,Liang:2021bde} and Taiji~\cite{Ruan:2018tsw}, while experiments that cover a similar range as BBO include DECIGO~\cite{Seto:2001qf,Kawamura:2006up}, AION~\cite{Badurina:2021rgt,Badurina:2019hst}, and AEDGE~\cite{Badurina:2021rgt,AEDGE:2019nxb}. We also note that current LIGO/VIRGO data has been used to search for a stochastic GW background~\cite{Romero-Rodriguez:2021aws}, but the results do not add any constraints to the parameter space that we show.

An overview of the method to obtain the bounds can be found in \cite{Thrane:2013oya}. 
We follow Appendix A of \cite{Schmitz:2020syl}, which calculates the strain sensitivity curves for these experiments explicitly using the methods in \cite{Thrane:2013oya}. The strain sensitivity curves and overlap reduction functions are used to calculate an effective GW background $\Omega_{\mathrm{eff}}$. The signal-to-noise (SNR) ratio for an expected stochastic GW spectrum $\Omega_{\mathrm{GW}}$ is then given by
\begin{equation}
\rho = \sqrt{n T} \left[ \int_{f_\mathrm{min}}^{f_\mathrm{max}} df \left( \frac{\Omega_{\mathrm{GW}}}{\Omega_{\mathrm{eff}}} \right)^2 \right]^{1/2}
\label{eq:SNR}
\end{equation}
where $T$ is the observational time and we take the integral over the entire bandwidth of the experiment. $n$ is 1 or 2 for auto-correlation (LISA, CE) and cross-correlation detections (BBO, ET), respectively. The spectrum $\Omega_\mathrm{GW}$ is function of initial perturbation amplitude $A$, scale $k_\star$, and Gaussian width $\sigma$, so for some choice of $\sigma$, we can solve for the parameters that give a desired SNR. In this paper, we present results with $\rho = 1$ and $T = 1\, \mathrm{yr}$ for each experiment and assume a perfect subtraction of foreground GWs from active sources. For a given $\sigma$, we can then use the one-to-one correspondence between $(k_\star,A)$ and  PBH parameters from Sec.~\ref{sec:BH_from_perturbations} to show the same bounds in $(m_\mathrm{BH}, \beta')$ space.

\begin{figure*}
\centering
\includegraphics[width=0.47\linewidth]{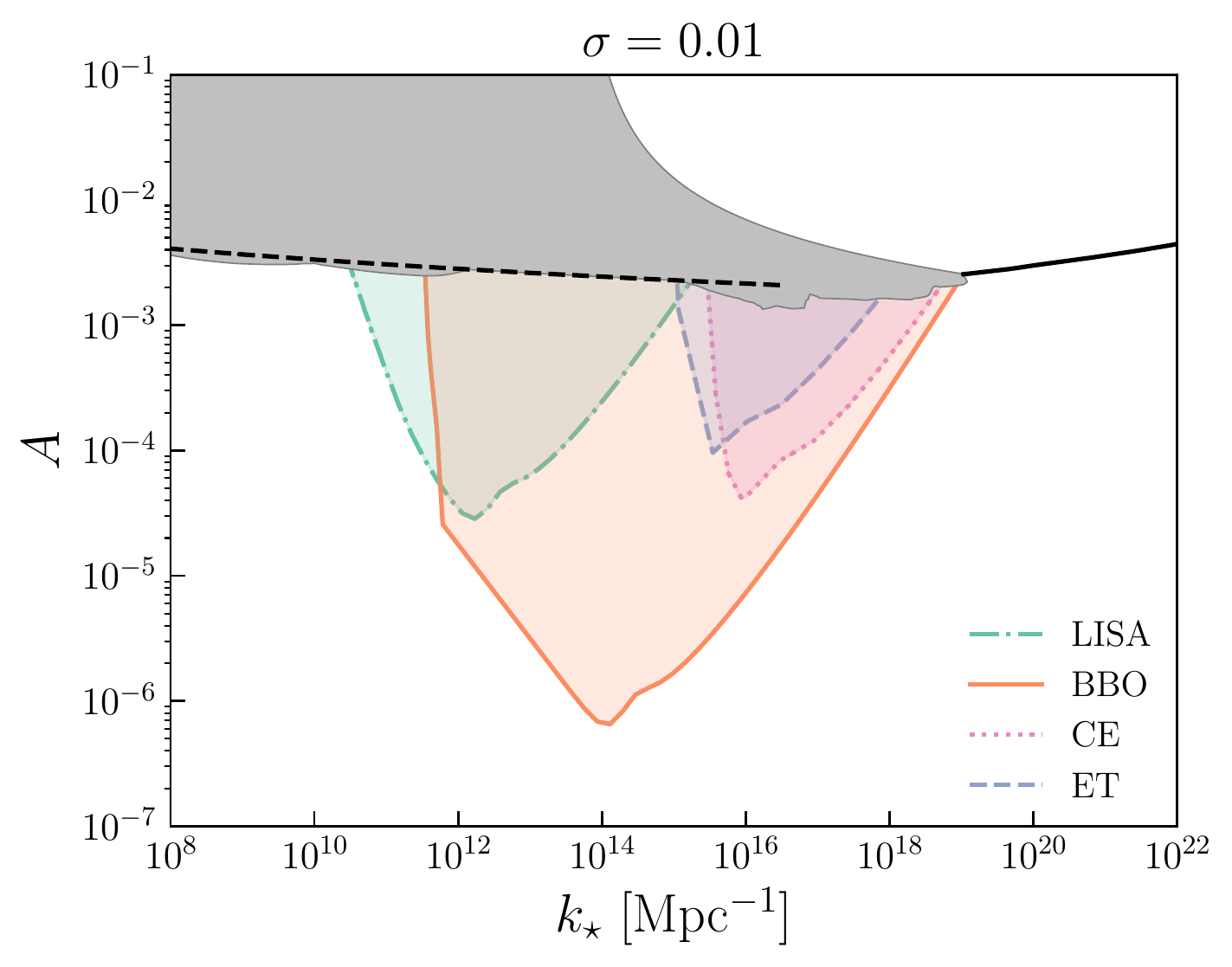}
\includegraphics[width=0.47\linewidth]{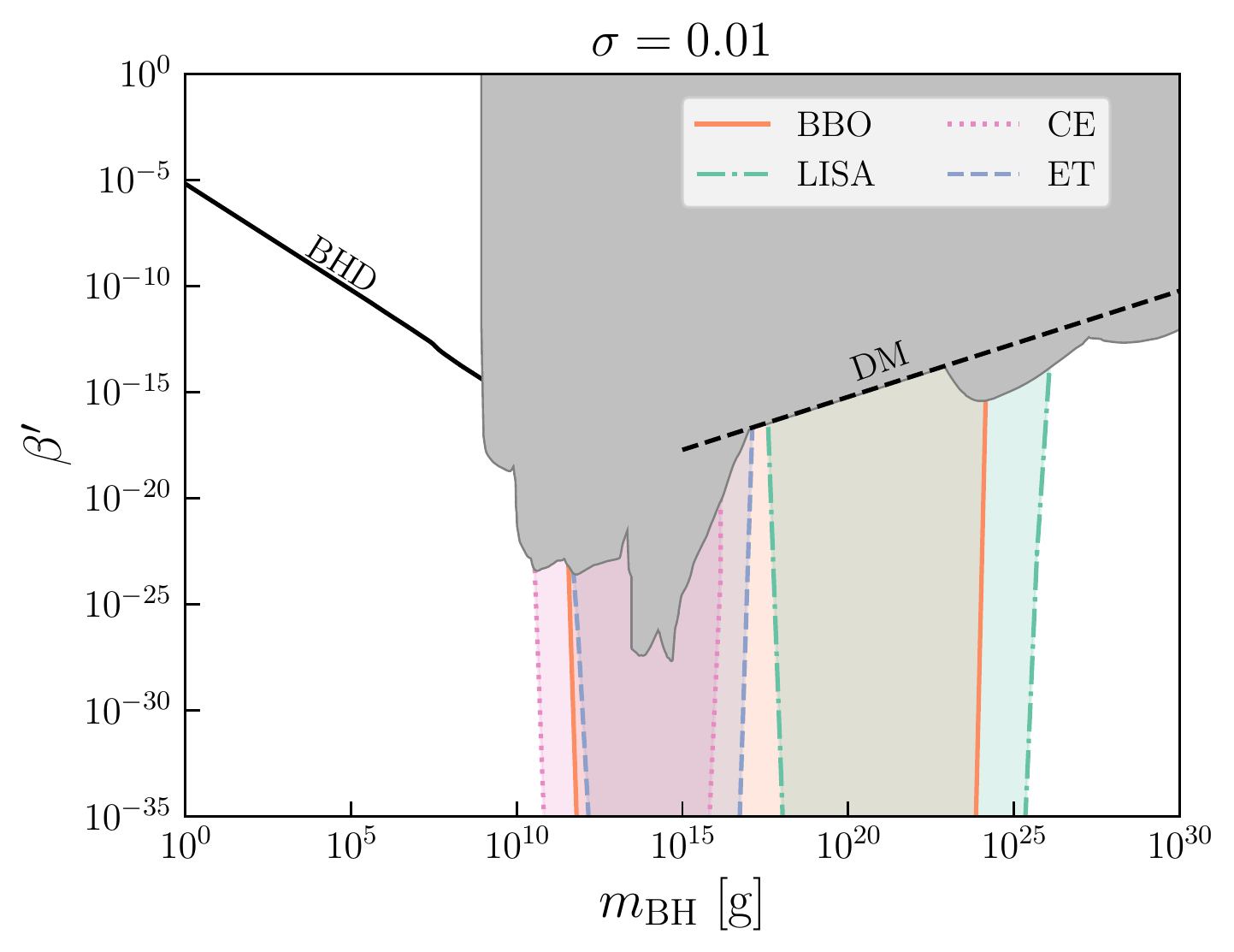} \\
\includegraphics[width=0.47\linewidth]{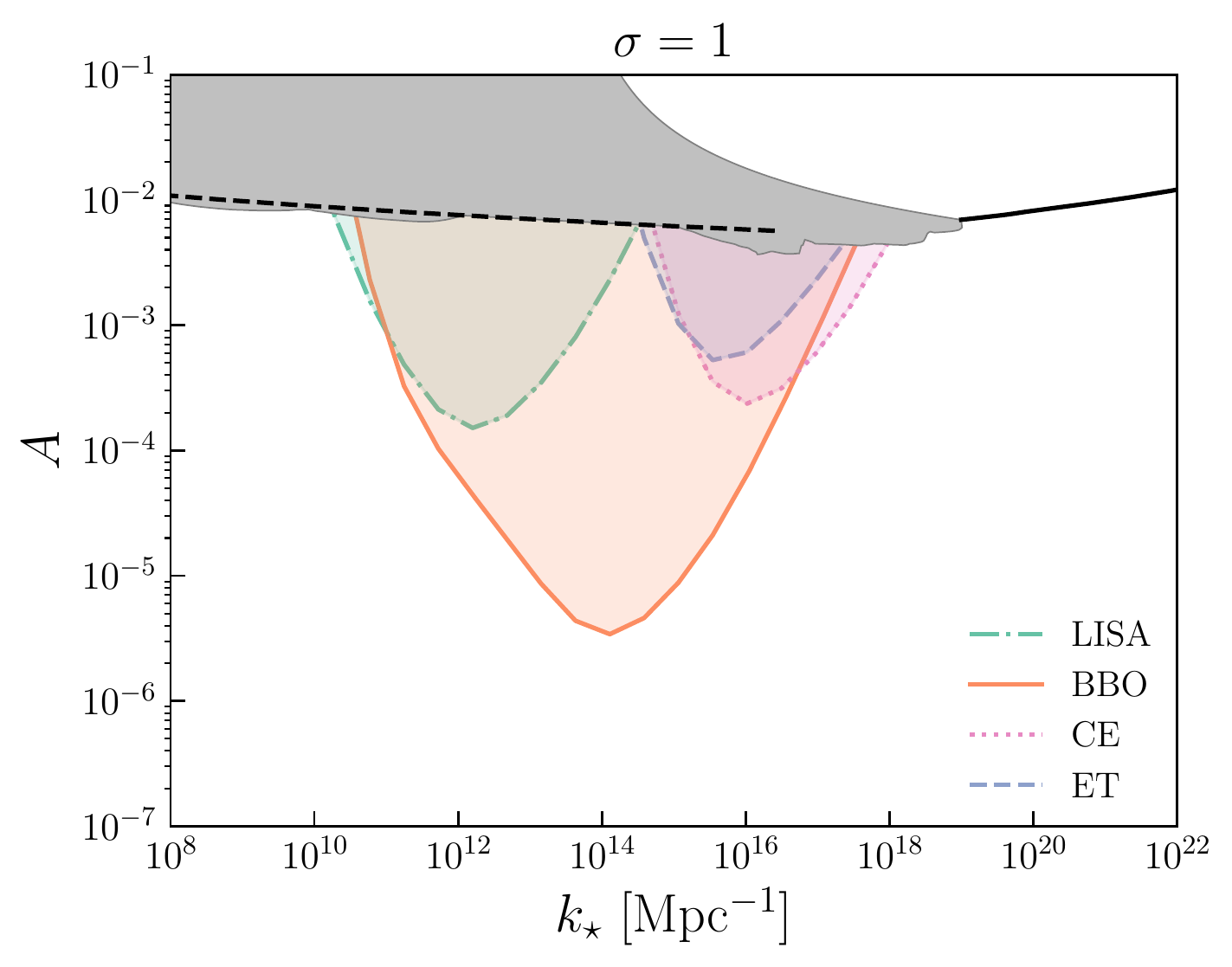}
\includegraphics[width=0.47\linewidth]{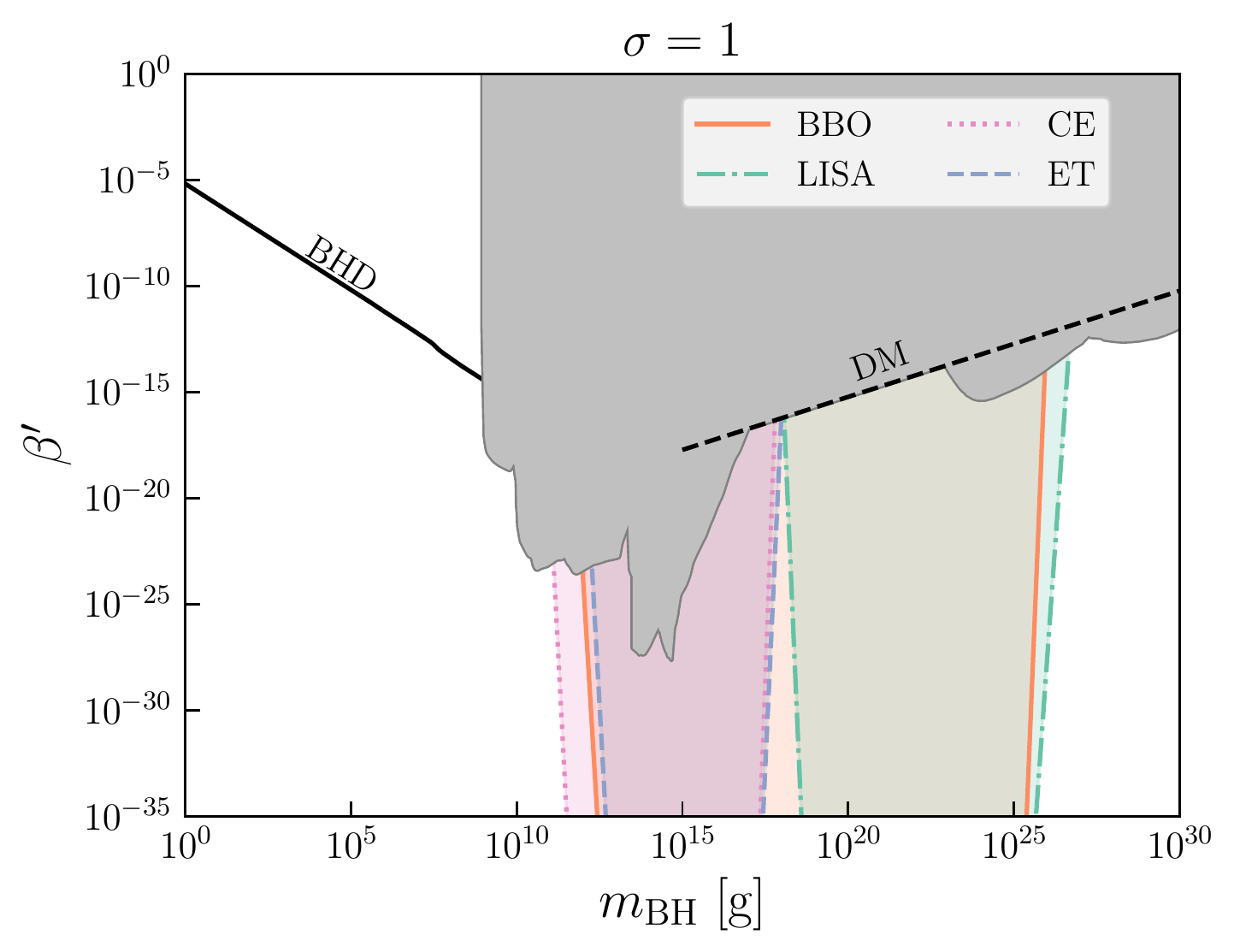} \\
\includegraphics[width=0.47\linewidth]{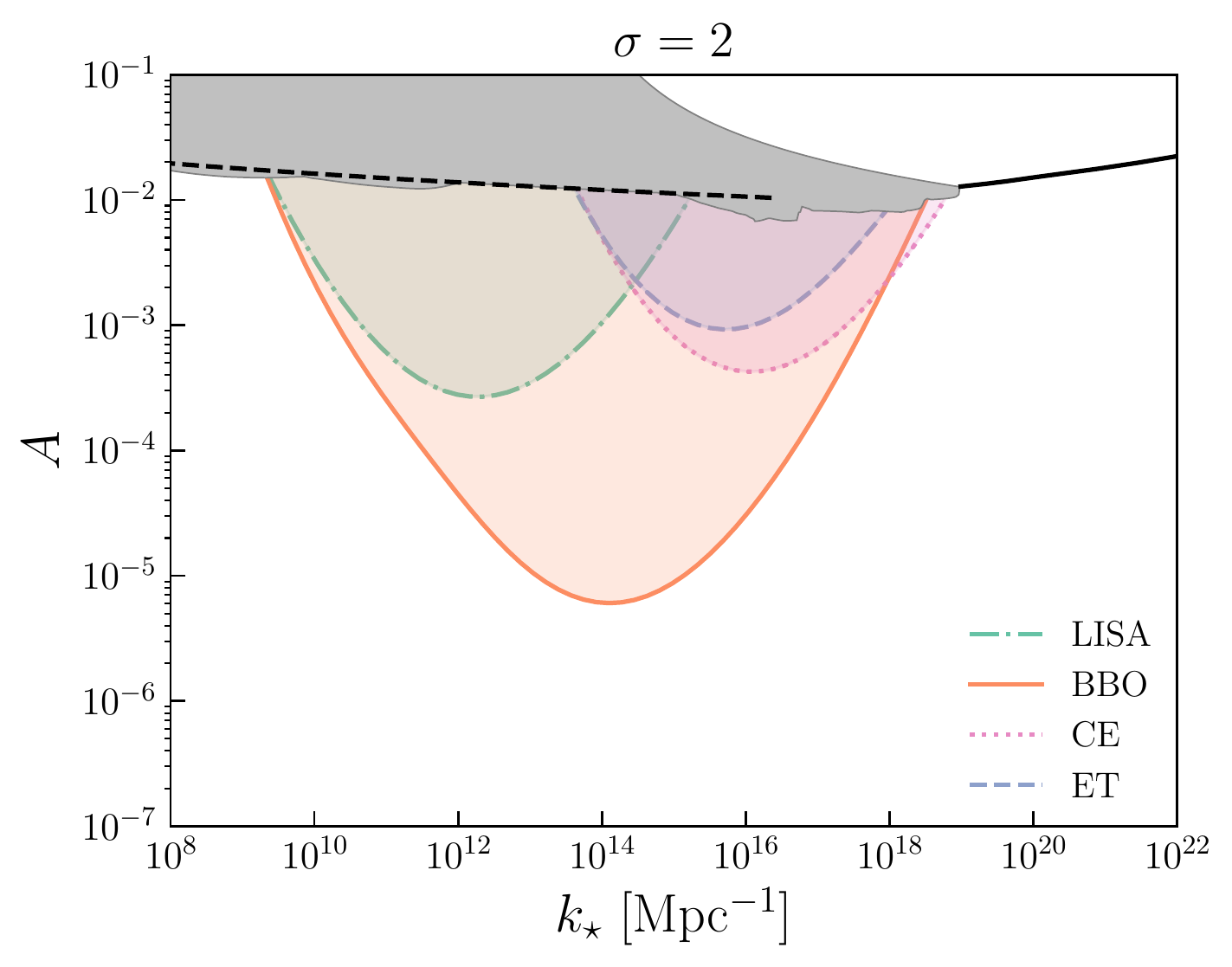}
\includegraphics[width=0.47\linewidth]{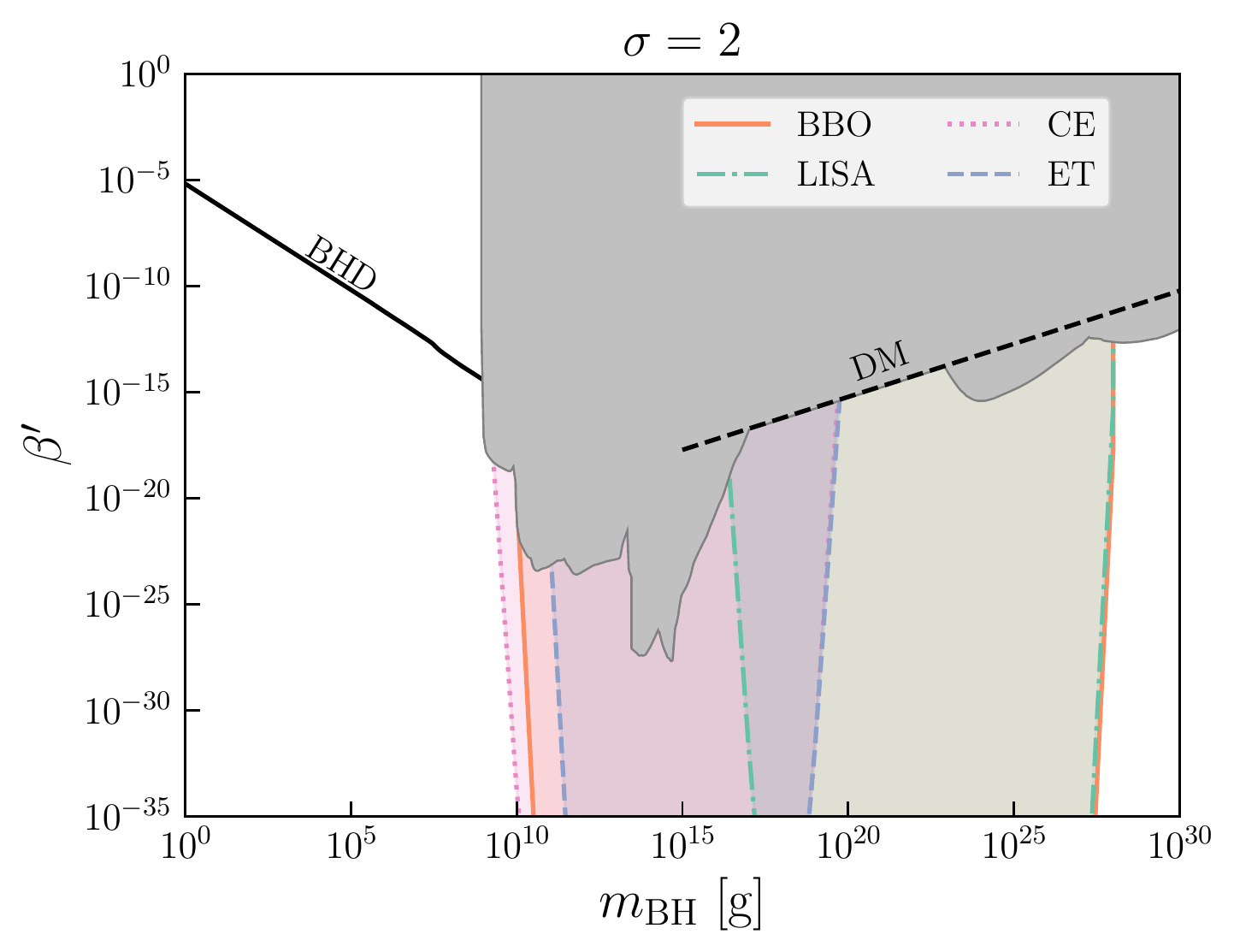}
\caption{Similar to Fig.~\ref{fig:monochromatic_bounds}, but assuming a Gaussian perturbation as in Eq.~\ref{eq:gaussian_perturbation}. The different rows are for different curvature perturbation widths $\sigma$.}
\label{fig:sigma_bounds}
\end{figure*}

\subsection{Results}

Our main results are shown in Fig.~\ref{fig:monochromatic_bounds} for a monochromatic perturbation and Fig.~\ref{fig:sigma_bounds} for a finite width Gaussian perturbation. The left panels show the existing bounds and experimental reach in the parameter space of primordial curvature perturbations, similar to Ref.~\cite{Inomata:2018epa}, while the right panels are in the parameter space for PBHs. In both panels, the gray shaded region are existing bounds on PBHs, as previously shown in Fig.~\ref{fig:BHD_regions}. Similar to Fig.~\ref{fig:BHD_regions}, we again indicate where PBHs comprise all of the DM (dashed line) and the lower boundary to the region where we have a black-hole dominated (BHD) era (solid line). These features appear roughly inverted in the left and right panel due to the relationship between $m_\mathrm{BH}$ and $k_\star$. We note again that the correspondence between $k_\star$ and $m_\mathrm{BH}$ is different if BHD occurs. If BHD never occurs, we have $k_\star \sim m_\mathrm{BH}^{-1/2}$, while if BHD does occur we instead have $k_\star \sim m_\mathrm{BH}^{-5/6} \beta^{-1/3}$. This is why the PBH bounds depend on both $A$ and $k_\star$ in the BHD era.

For black holes that evaporate after BBN ($m_\mathrm{BH} > 10^9 \ \mathrm{g}$), gravitational wave interferometers are sensitive to the stochastic GW spectrum induced at second order from the primordial curvature perturbation. In this mass range, we see the broadest level of sensitivity for the largest curvature perturbation width ($\sigma=2$), with the reach extending from $10^9 - 10^{28}$ g when all experiments are included. This is because the GW spectrum extends over a much wider frequency range, as seen in Fig.~\ref{fig:different_sigma}. The reach curves thus shrink with smaller $\sigma$ in  Fig.~\ref{fig:sigma_bounds}. However, once $\sigma$ decreases below $\sigma \sim 0.01$, the reach actually improves again at small PBH mass, which can be seen in the sensitivities for the monochromatic perturbation in Fig.~\ref{fig:monochromatic_bounds}. This is because the $k^2$ infrared tail grows as $\sigma$ decreases below $\sigma \sim 0.01$, as discussed in Sec~\ref{subsec:primordial_perturbation}.

In the region of parameter space where the black holes are short-lived, the spectrum of stochastic GWs produced during or at the end of the BHD era is highly uncertain since density perturbations become nonlinear. At present, we do not find a detectable signal from any of the sources discussed in Secs.~\ref{sec:GWs_from_BHD}-\ref{sec:GWs_from_reheat}. However, more numerical work should be done to investigate what happens in a BHD era as density perturbations become non-linear and whether there are observable GW sources from the transition of the BHD to radiation era.

\section{Conclusions}
\label{sec:conc}
In this paper, we have investigated how generation GW interferometers can probe the parameter space of primordial black holes in the mass range of 10~g -- $10^{28}$~g. 
%The experimental sensitivity extends well beyond existing bounds, including a previously unbounded parameter space of light black holes that dominate the universe before evaporation. 
Assuming that the black holes form from some generic primordial curvature perturbation at small scales generated by inflation, there are several possible stochastic GW signals, which we calculate as a function of the initial abundance and mean black hole mass. Second order gravitational waves from the primordial curvature perturbation can be used to probe primordial black hole masses in the range $~10^9-10^{28}$ g down to very small $\beta'$, depending on the experiment and curvature perturbation width. For black hole masses below $10^9$ g, there are additional possible sources of GWs if $\beta'$ is sufficiently large enough, such that there is an early black hole dominated era. We considered second order GWs generated from the black hole fluid curvature perturbation itself, as well as GWs produced from the turbulent fluid at the end of the BHD era. Based on our estimates, we did not find these to give rise to observable GWs.
In addition, there may be GWs produced during the nonlinear evolution of the BHD era, from Hawking evaporation itself, mergers, or from a sharp transition from black hole dominated to radiation eras. Treating the BHD era properly and including all of these sources requires a numerical treatment.

Finally, we also presented results in terms of the curvature perturbation parameters and see that second order GWs allow us to probe perturbations at quite small scales. The stochastic GW spectrum is thus a promising way to probe inflationary scenarios giving rise to both short-lived black holes, as well as long-lived black holes comprising some fraction of the dark matter today.

\acknowledgments

We thank David Langlois, Theodoros Papanikolaou, and Vincent Vennin for helpful discussions, and Vincent Vennin for valuable feedback on a draft of this work. JK is supported by the Department of Energy under grants DE-SC0019195 and DE-SC0009919. TL is supported by the Department of Energy under grant DE-SC0019195 and a UC Hellman fellowship. EV is supported by a Sloan Scholar Fellowship.

\appendix

\section{GWs in BHD era}
\label{sec:nonlinear_cutoff}

\begin{figure*}[t]
    \centering
    \includegraphics[width=0.47\linewidth]{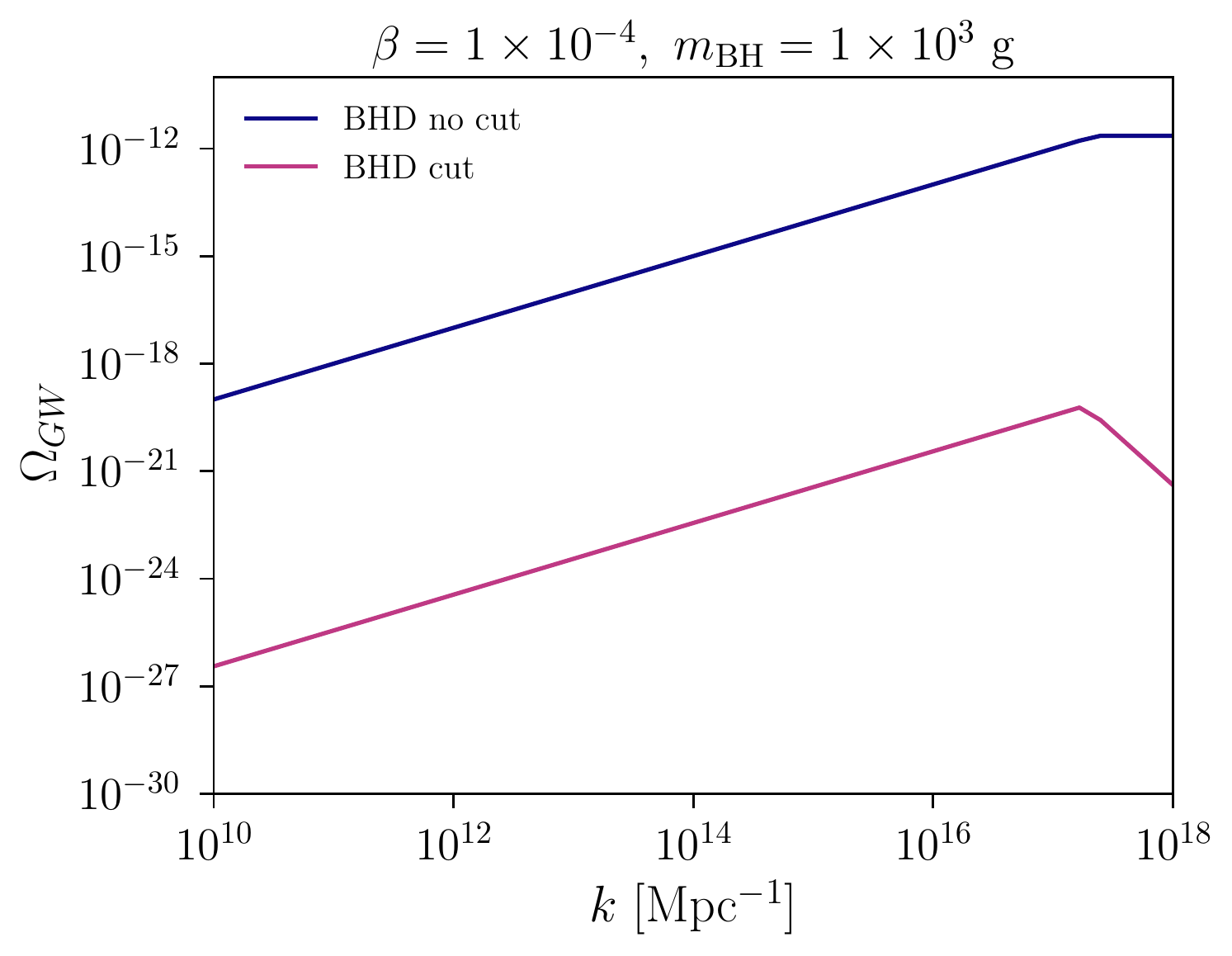}
    \includegraphics[width=0.47\linewidth]{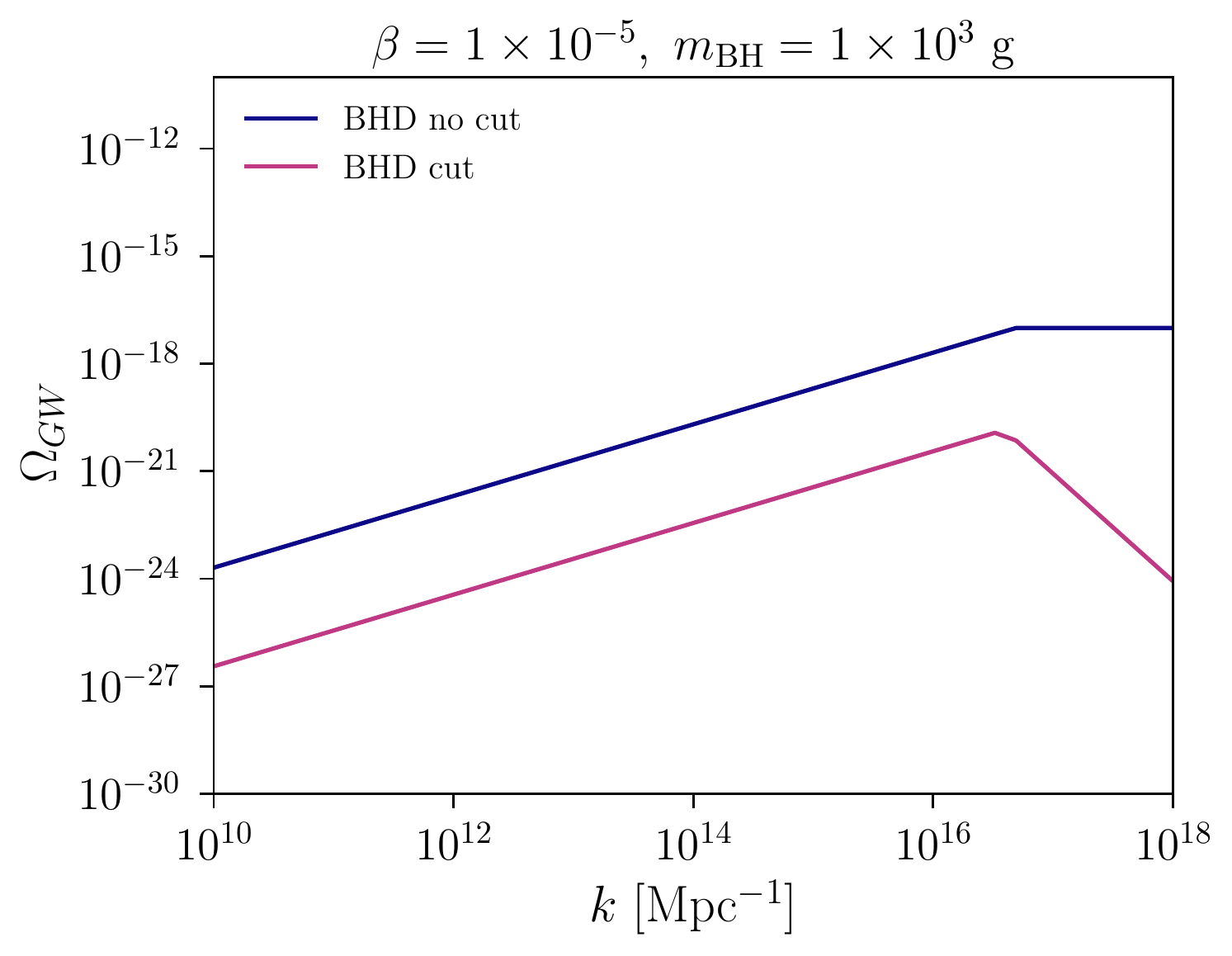}
\caption{Comparison of GW spectra today from the BHD era and turbulence for $\MBH = 10^3 \mathrm{g}, \sigma=1$ and two values of $\beta$. The blue line is the pure BHD era result (\ref{eq:omega_bhd}) without any modification to account for the fact that the density perturbations become non-linear. The magenta line is our conservative lower bound of the spectrum, which includes a cut-off when the density perturbations become non-linear. This introduces an additional suppression factor (\ref{eq:bhd_Rfactor}) to (\ref{eq:omega_bhd}).}
    \label{fig:turbulence_bhd_comparisons}
\end{figure*}

In this appendix, we briefly describe how we obtained \eqref{eq:bhd_Rfactor}, which we use to obtain a rough lower bound on second-order GWs sourced by scalar perturbations during the BHD era.

To obtain \eqref{eq:bhd_Rfactor}, we imposed $k$-dependent time cutoff on density perturbations. To apply this cutoff, we must re-calculate the quantity $I$, which now becomes
\begin{align}
    \label{eq:I_cutoff}
    I(u,v,x) = \int_{x_d}^x d\bar{x} \frac{a(\bar{x})}{a(x)} k G_k(x, \bar{x}) F_k (u, v, \bar{x}) \\ \nonumber \times \Theta(x_{\mathrm{cut}} (v k) - \bar{x}) \Theta(x_{\mathrm{cut}} (u k) - \bar{x}).
\end{align}
where $G_k$ and $F_k$ are Green's functions defined in \cite{Papanikolaou:2020}, $x = k \eta_\mathrm{RH}$, and $x_\mathrm{cut} = k \eta_\mathrm{cut}$ is given by
\begin{align}
    \label{eq:xcut_definition}
    & x_{\mathrm{cut}} (v k) = \nonumber \\
    & x \sqrt{ \mathrm{Min} \Bigg[ \frac{a_\mathrm{BHD}}{a_\mathrm{RH}} \times
\begin{cases}
 \sqrt{\frac{3\pi}{2}} \left( \frac{\mathcal{H}_\mathrm{d}}{v k}\right)^2\left(\frac{k_\mathrm{UV}}{v k}\right)^{3/2} & vk < \mathcal{H}_\mathrm{d} \\
 \sqrt{\frac{3\pi}{2}} \left(\frac{k_\mathrm{UV}}{v k}\right)^{3/2} & v k > \mathcal{H}_\mathrm{d} \\
\end{cases}, 1 \Bigg]}.
\end{align}
This comes from applying \eqref{eq:nonlinear_cut_redshift} to the density perturbation $vk$. Evaluating $I^2$ as in Appendix B of \cite{Papanikolaou:2020} gives the leading term
\begin{equation}
I^2 \sim \frac{100}{9} \frac{(x_{\mathrm{cut}}(vk))^4}{x^4},
\end{equation}
which is similar to the result in \cite{Papanikolaou:2020}, but with an additional suppression from the cutoff on nonlinear perturbations. 
Since \eqref{eq:I_cutoff} has two step functions we should have $\mathrm{Min}[x_{\mathrm{cut}}(vk),x_{\mathrm{cut}}(uk)]$, but the integrand always peaks near $u \sim v$ so we take both step functions to be the same.

We now evaluate the tensor spectrum with our modification to $I^2$. Here we provide simple estimates in the large and small $k$ limits, similar to \cite{Papanikolaou:2020}. Since the integrand (aside from the $I^2$ factor) is peaked in $u, v$, we will treat  $I^2$ as constant about the peak. Then the effect of the nonlinear cutoff is just to introduce a factor of $(x_{\mathrm{cut}}(vk)/x)^4$, evaluated at the peak value of $v$. For $k \gg k_\mathrm{BHD}$, the integrand is peaked at $u = v = 1$, so we can approximate $x_{\mathrm{cut}} (v k) \approx x_{\mathrm{cut}} (k)$. For $k \ll k_\mathrm{BHD}$, the integral is peaked at $u \sim v \sim \frac{6 k_\mathrm{BHD}}{k}$ which gives  $x_{\mathrm{cut}} (v k) \sim x_{\mathrm{cut}} (6 k_{\mathrm{BHD}})$. Using \eqref{eq:xcut_definition} in these limits then gives \eqref{eq:bhd_Rfactor}, where we glued the solutions in the two limits together to obtain a continuous function. We have checked numerically that including the $u,v$ dependence  only leads to $O(\mathrm{few})$ deviations from our analytic estimates.

Fig.~\ref{fig:turbulence_bhd_comparisons} shows a comparison of the GW spectrum without the nonlinear cutoff, \eqref{eq:omega_bhd}, and with the cutoff. With the nonlinear cutoff, our estimate of this GW signal is not detectable.

\bibliography{gw}

\end{document}